\documentclass[aps,prl,reprint,twocolumn,showpacs,amsmath,amssymb]{revtex4-1}

\usepackage{graphicx}
\usepackage{amssymb}
\usepackage{amsmath}
\usepackage{dcolumn}
\usepackage{bm}
\usepackage{mathrsfs}
\usepackage{color}
\usepackage{hyperref}

\begin{document}

\title{Heralded Preparation and Readout of Entangled Phonons\\ in a Photonic Crystal Cavity}

\author{Hugo Flayac}
\author{Vincenzo Savona}
\affiliation{Institute of Theoretical Physics, \'{E}cole Polytechnique F\'{e}d\'{e}rale de Lausanne EPFL, CH-1015 Lausanne, Switzerland}

\begin{abstract}
We propose a realistic protocol for the preparation and readout of mechanical Bell states in an optomechanical system. The proposal relies on parameters characterizing a photonic crystal cavity mode, coupled to two localized flexural modes of the structure, but equally applies to other optomechanical systems in the same parameter range. The nonclassical states are heralded via optical postselection and revealed in specific interference patterns characterizing the emission at the cavity frequency.
\end{abstract}
\maketitle

Entanglement is the most distinctive feature of quantum mechanics \cite{Horodecki2009}, and represents the essential resource for quantum information processing \cite{Kimble2008}. One of the most striking predictions of quantum theory is that entanglement may exist up to the macroscopic realm. In reality, the entanglement of systems with several degrees of freedom is countered by fast interaction with the environment, and the deepest essence of such {\em decoherence} process has not yet been unraveled. Experiments along this line might clarify the mechanisms underlying decoherence, or even hint at a fundamental restriction of the quantum predictions, that would define a quantum-classical boundary. Within a more applied perspective, entanglement at the macroscopic scale -- especially in integrated solid-state microstructures -- will be required for quantum information storage and communication \cite{Kimble2008,Duan2001}. Recently, several groundbreaking experiments have succeeded in producing entanglement of macroscopic degrees of freedom, involving ultracold atoms \cite{Choi2008,Chou2005,Laurat2007}, photons \cite{Bruno2013,Lvovsky2013}, and collective electronic excitations \cite{Usmani2012}.

A promising route to macroscopic entanglement is optomechanics \cite{Aspelmeyer2013,Favero2009,Kippenberg2008,Marquardt2009}. In an optomechanical system, electromagnetic modes of an optical cavity are coupled to one or more mechanical oscillators via radiation pressure and mechanical backaction. Several theoretical studies have suggested the generation nonclassical mechanical states \cite{Pirandola2006,Zhou2011,Vanner2013,Ghobadi2014,Galland2014}, and more specifically entangled states of two mechanical modes \cite{Akram2013,Borkje2011,Ge2013,Hartmann2008,Ludwig2010,Pinard2005,Szorkovszky2014,Tan2013a,Xu2013}. Recently, an entangled state of optical phonons of two distant bulk diamond crystals has been produced and detected \cite{Lee2011}.

Here, we demonstrate a full protocol for heralded entanglement generation and readout of two mechanical modes in a realistic optomechanical system. The protocol is specifically studied having in mind two localized flexural modes of a $L3$ photonic crystal cavity, for which significant optomechanical coupling strengths and close-valued mechanical frequencies have been demonstrated \cite{Gavartin2011}. Silicon photonic crystal cavities are increasingly considered as an ideal building block of an integrated technological platform for quantum photonics, and the present protocol might indicate a way to the generation and storage of quantum information. It can however equally be applied to other optomechanical systems for which the state-of-the-art parameters, such as mechanical quality factor and optomechanical coupling, are steadily improving \cite{Ding2011,Chan2011,Chan2012,Gavartin2013,Nguyen2013,Safavi-Naeini2013,Safavi-Naeini2014}.

\begin{figure}[ht!]
\includegraphics[width=0.47\textwidth,clip]{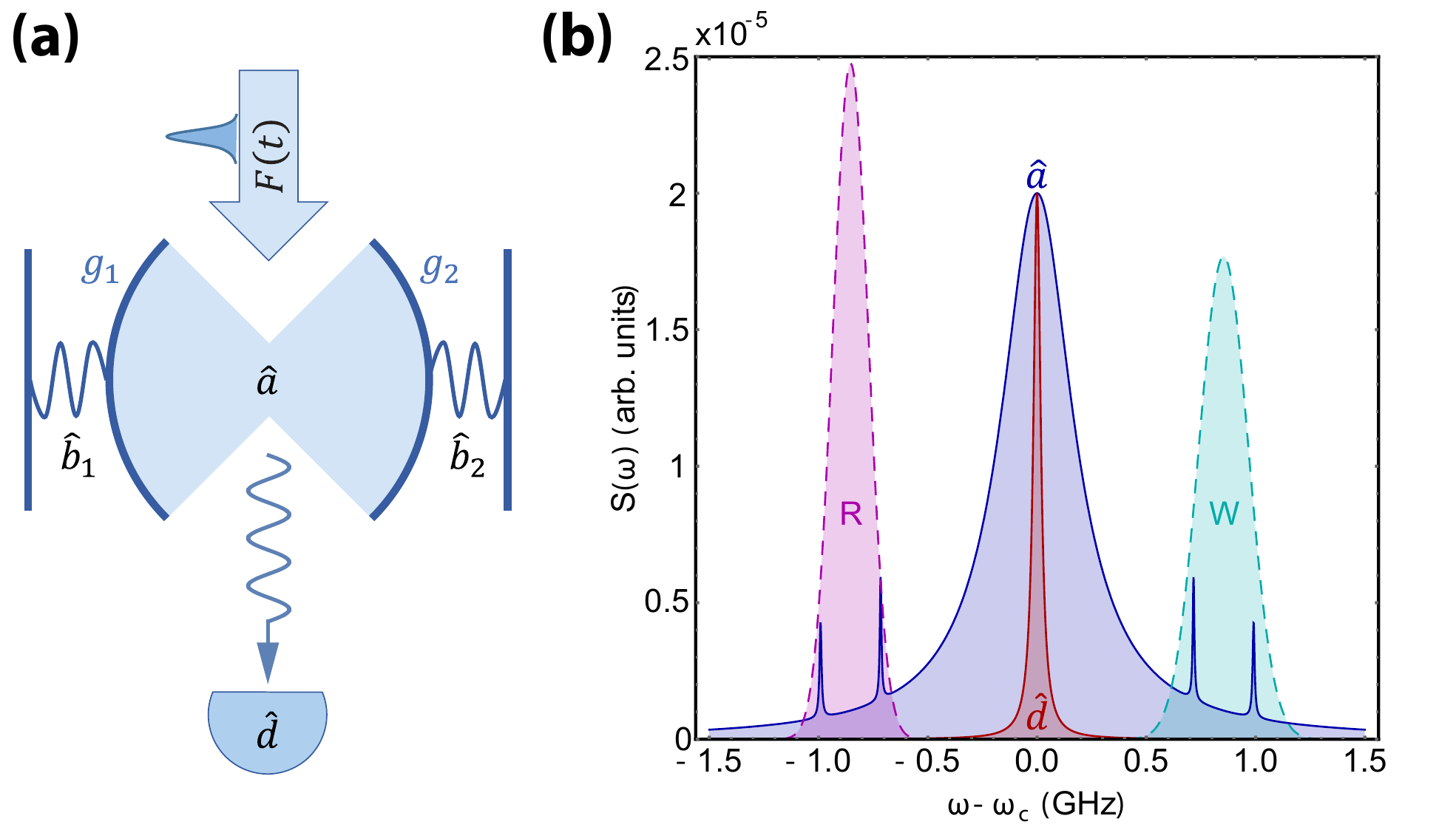}\\
\caption{(a) Schematic representation of the system: Single mode cavity coupled to 2 mechanical modes and driven by pulsed excitations. The emission is collected through a narrow band detector $d$. (b) Cavity (blue) and detector (red) power spectrum. The cyan (magenta) curves show the write (readout) pulse (not on scale).}
\label{Fig1}
\end{figure}

We consider the system sketched in Fig.\ref{Fig1}(a) composed of an optical cavity with a mode ($\hat{a}$ operator) resonant at frequency $\omega_c$, optomechanically coupled to two harmonic mechanical modes ($\hat{b}_{1,2}$  operators) of resonant frequencies $\Omega_{1,2}$. The system Hamiltonian reads
\begin{eqnarray}\label{H}
\nonumber {\cal{\hat H}}_s &=&\hbar\omega_c\hat a^\dag\hat a+ \sum\limits_{j = 1,2}\left[ {{\hbar\Omega _j}\hat b_j^\dag {{\hat b}_j} - g_j{{\hat a}^\dag }\hat a\left( {\hat b_j^\dag  + {{\hat b}_j}} \right)}\right]\\
      &+& F\left( t \right){\hat a^\dag } + {F^*}\left( t \right)\hat a\,.
\end{eqnarray}
Here, $g_{1,2}$ are the single-photon optomechanical coupling constants, and the cavity is driven by a classical field $F(t)$. We assume parameters in the range of those measured in Ref.\cite{Gavartin2011}, i.e. $\Omega_1/2\pi=700~\mbox{MHz}$, $\Omega_2/2\pi=980~\mbox{MHz}$, $g_1/2\pi=72~\mbox{kHz}$, $g_2/2\pi=84~\mbox{kHz}$, and mechanical loss rates $\gamma_1/2\pi=4.4~\mbox{MHz}$ and $\gamma_2/2\pi=5.4~\mbox{MHz}$. In that work, both optical and mechanical modes were thoroughly characterized both theoretically and experimentally, while a general model of the optomechanical coupling $g$ in photonic crystals has been derived in Ref. \cite{Safavi-Naeini2010}. Selective Stokes and anti-Stokes driving however requires the system to be in the well resolved sideband limit. Hence, we propose the use of a last-generation silicon $L3$ photonic crystal cavity, for which quality factors up to two millions were recently predicted and measured \cite{Lai2014,Minkov2014}. We therefore assume $\omega_c/2\pi=194~\mbox{THz}$, corresponding to $\lambda=1550~\mbox{nm}$, and an optical loss rate $\kappa/2\pi=200~\mbox{MHz}$ i.e. a quality factor $Q\simeq10^6$. This value, conservative if compared to those measured in Ref. \cite{Lai2014}, should also account for additional leakage caused by a near-field optical detection scheme \cite{Chan2012}. Fig.\ref{Fig1}(b) shows the computed optical spectrum (blue curve) consisting of the main cavity resonance, and two pairs of sidebands associated to Stokes and anti-Stokes processes.

Before presenting the details of the protocol, let us illustrate its general principle through a simplified analysis based on pure states: We denote by $|n_c,n_1,n_2\rangle$ the number state with $n_c$ photons and, respectively, $n_1$ and $n_2$ quanta in the first and second mechanical mode. We assume that the system is initially in its vacuum state $|\psi_0\rangle=|000\rangle$. The writing phase consists in driving the cavity with a weak optical pulse resonant with the two sidebands, as sketched in Fig.\ref{Fig1}(b). Two Stokes processes will take place, each leaving a phonon in a mechanical mode and a photon in the cavity mode. For a weak enough pulse, the dominant terms in the final state will be the vacuum, followed by the state with one photon, i.e. $|\psi_1\rangle=\alpha_1|000\rangle+\beta_1|1_c\rangle\left(|10\rangle+e^{i\phi_s}|01\rangle\right)+\ldots$,
where $\phi_s$ is a phase, that characterizes the coherent superposition of the two Raman processes. After the writing pulse has fully decayed, the heralding is performed by detecting the emission spectrally filtered at $\omega_c$. If a photon is detected, then the vacuum component is projected out giving the state $|\psi_2\rangle=\beta_2|0_c\rangle\left(|10\rangle+e^{i\phi(t)}|01\rangle\right)+\ldots$,
where the phase $\phi(t)=\phi_s-(\Omega_2-\Omega_1)t$ evolves due to the nondegeneracy of the two mechanical modes. In the limit where higher-occupation states have negligible amplitudes, this is a maximally entangled state of the two mechanical modes. Conditioned to the detection of a heralding photon, a readout is performed by driving the system with a second optical pulse resonant with the two anti-Stokes sidebands, as sketched in Fig. \ref{Fig1}(b) (red curve). As for the writing phase, a coherent superposition of two anti-Stokes processes will bring the system to the state $|\psi_3\rangle=\alpha_3(1+e^{i\phi^\prime(t)})|1_c00\rangle+\beta_3|0_c\rangle(|10\rangle+e^{i\phi(t)}|01\rangle)+\ldots$,
where $\phi^\prime(t)=\phi(t)+\phi_{as}$ accounts, as above, for the relative phase of the two anti-Stokes processes. If emission at $\omega_c$ is again detected, after the readout pulse decayed, the emitted intensity will be $I\propto|1+e^{i\phi^\prime(t)}|^2$, i.e. it will display a full-contrast interference pattern as a function of the photon detection time. If instead one has a fully separable state $|\psi_2\rangle=\beta_2|0_c\rangle\left(|0\rangle+e^{i\phi_1}|1\rangle\right)\left(|0\rangle+e^{i\phi_2}|1\rangle\right)$, then the visibility will be limited to $50\%$ \cite{Supplemental}.

This protocol follows in the footsteps of the procedure proposed in Refs.\cite{Vanner2013,Galland2014} to produce a mechanical Fock state. Similarly to other schemes studied in the past, \cite{Borkje2011,Lee2011} it is based on the original DLCZ proposal \cite{Duan2001} for entangling atomic ensembles. An important distinctive feature of the present proposal however, is that the light emitted by the two Stokes processes must not be mixed in a beam-splitter to erase the which-path information, as the two processes emit directly into the same mode.

The interference pattern in the emitted light is not in itself an entanglement witness. In particular, a similar interference pattern might arise from classical field amplitudes, provided they have the right mutual phase relation. However -- similarly to seminal quantum optics experiments \cite{Mandel1999,Zou1991} -- an interference pattern, combined with the knowledge that the average mode occupation fulfils $\langle \hat n\rangle\ll 1$, is a solid indication of the occurrence of a nonclassical state. Alternatively, a full quantum tomography of the mechanical state might in principle be carried out by sending a readout pulse at $\omega_c$ and detecting Raman photons at the two Stokes sidebands. The very low readout signal expected for this process however would pose a severe challenge for correlation measurements, highlighting the advantage of the simple readout scheme suggested here.

\emph{The Model}.---
The model considered here is based on Hamiltonian (\ref{H}) where the driving field is characterized by two pulsed excitations $F(t) = F_W(t) + F_R(t)$ where
\begin{equation}
F_j(t) = A_j{e^{ - \left( {t - {t_j}} \right)^2/\sigma_j^2}}{e^{ - i{\omega_j}t}},{\rm{}}j=W,R
\end{equation}
respectively for the write and readout procedures. Here, $\omega_j$, $t_j$, $\sigma_j$, and $A_j$ are the frequency, time, bandwidth, and amplitude, respectively for the two pulses.

The protocol relies on photon detection spectrally filtered at the cavity frequency $\omega_c$. A computationally very effective way to model spectral filtering consists in introducing an additional mode, representing the degrees of freedom of the narrow-band detector \cite{Valle2012}. Its Hamiltonian is expressed as ${\cal{\hat H}}_d=\hbar{\omega_d}{\hat d^\dag }\hat d$, and the frequency $\omega_d$ and damping rate $\kappa_d$ define the frequency and passband of the filter [see red curve in Fig. 1(b)]. The global Hamiltonian reads ${\cal{\hat H}}={\cal{\hat H}}_s+{\cal{\hat H}}_d$.

In presence of coupling to the environment, the system dynamics is governed by the master equation for the density matrix, which in the Lindblad form \cite{Gardiner2004} reads
\begin{eqnarray}
\nonumber \frac{{d\hat \rho }}{{dt}} = &-& i\left[ {{\cal{\hat H}},\hat \rho } \right] - \frac{{{\kappa}}}{2}{\cal{D}}\left[ {\hat a} \right]\hat \rho  - \frac{{{\kappa _d}}}{2}{\cal{D}}\left[ {\hat d} \right]\hat \rho + \frac{{{\zeta }}}{2}{{\cal{D}}}\left[ \hat a, \hat d \right]\hat \rho \\
\label{rhot}
 &-& {{\bar n}_{th}}\sum\limits_j {\frac{\gamma_j }{2} {{\cal{D}}\left[ {{{\hat b}_j}} \right]\hat \rho  - \left( {{{\bar n}_{th}} + 1} \right)\frac{\gamma_j }{2}{\cal{D}}\left[ {\hat b_j^\dag } \right]\hat \rho }}\,.
\end{eqnarray}
Here, ${\cal{D}}\left[ {\hat o} \right]\hat \rho  = {{\hat o}^\dag }\hat o\hat \rho  + \hat \rho {{\hat o}^\dag }\hat o - 2\hat o\hat \rho {{\hat o}^\dag }$ describe the coupling to the environment at rates $\kappa$, $\kappa_d$ and $\gamma_j$ for the cavity, detector and mechanical modes respectively. $\bar n_{th}=[\exp(\Omega/k_BT)-1]^{-1}$ is the mean thermal phonon number associated to a mechanical frequency $\Omega$. Contrarily to Ref.\cite{Valle2012}, we assume here a dissipative, thus irreversible, cavity-to-detector coupling, described by ${\cal{D}}[ {\hat a,\hat d} ]\hat \rho  = [ {\hat a\hat \rho ,{{\hat d}^\dag }} ] + [ {\hat d\hat \rho ,{{\hat a}^\dag }} ]$. This approach prevents any unwanted coherent oscillations between the two modes and therefore allows to consider an arbitrary coupling strength $\zeta$. From now on, we shall consider $\omega_d=\omega_c$ and $\kappa_d=\zeta=0.1\kappa$.

Equation (\ref{rhot}) was solved numerically: For the write phase, a finite dimensional Hilbert space, restricted to a maximum number of quanta for each mode, was assumed. The readout phase is characterized by a significantly stronger driving field. Hence, each field was decomposed into the sum of a classical and a quantum fluctuation component. For the cavity field we define $\hat a=\langle\hat a\rangle+\delta\hat a$, and analogously for the other fields \cite{Supplemental}.

\begin{figure}[ht]
\includegraphics[width=0.47\textwidth,clip]{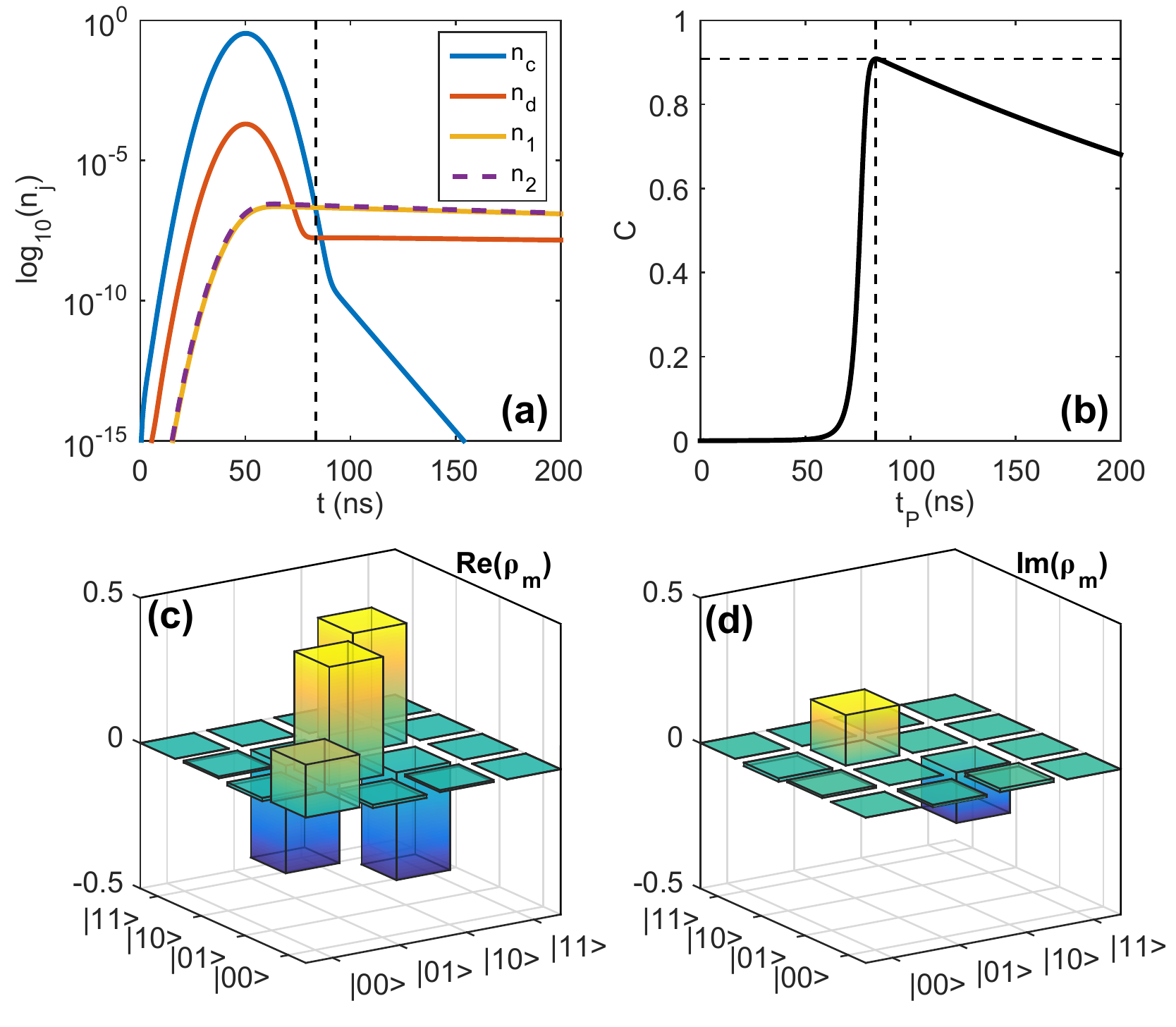}\\
\caption{(a) Average occupations of the cavity (blue line), detector (red line) and mechanical (yellow and purple lines) modes as a function of time. (b) Concurrence $C$ of ${\hat{\rho}}_m$ as a function of $t_P=t$. The vertical dashed line in (a) and (b) marks the concurrence maximum, occurring at $t_P=83$ ns. (c) Real and (d) imaginary parts of $\hat{\rho}_m$, computed at the concurrence maximum.}
\label{Fig2}
\end{figure}

\emph{Write}.---
The write step starts with the arrival of the first pulse, $F_W(t)$. We set the parameters to $A_W=2.5\kappa$ -- low enough to ensure a negligible two photon occupation -- $t_W=50$ ns, $\sigma_W=12.5$ ns and $\omega_W=\omega_c+(\Omega_1+\Omega_2)/2$ which allows the simultaneous excitation of both Stokes sidebands.  The cavity, mechanical, and detector average occupations $n_c(t)={\langle{{a^\dag }a}\rangle}$, $n_b{_j}(t)={\langle{{b_j^\dag }b_j}\rangle}$ and $n_d(t)={\langle{{d^\dag }d}\rangle}$ are plotted in Fig.\ref{Fig2}(a), assuming that the heralding photon has not been detected. The plot shows that, after the writing pulse, a small average occupation of the mechanical modes is produced through a Stokes Raman process, and then decays as a result of mechanical damping.

The second step of the write procedure consists in heralding the formation of an entangled state, through the detection at time $t_P>t_W$ of a photon at the cavity frequency $\omega_c$. This projects the system onto the subspace where one photon is present in the detector mode, i.e.
\begin{eqnarray}
{{\hat \rho }_P} &=& \frac{{\hat{\cal{P}}_1\hat \rho \left( t_P \right)\hat{\cal{P}}_1^{-1}}}{{{\rm{Tr}}\left[ {\hat {\cal{P}}_1\hat \rho \left( t_P \right)} \right]}},\label{rhop}\\
{{\hat {\cal{P}}}_{1}} &=& {\mathbb{\hat{I}}_{\hat a}} \otimes {\mathbb{\hat{I}}_{{\hat{b}_1}}} \otimes {\mathbb{\hat{I}}_{{\hat{b}_2}}} \otimes {\left| 1_{\hat{d}} \right\rangle}\langle 1_{\hat{d}}{|}
\end{eqnarray}
where $\mathbb{\hat{I}}_{\hat a}$ and $\mathbb{\hat{I}}_{\hat{b}_{j}}$ are the identity operators acting on the cavity and mirror subspaces respectively. $\left| 1_{\hat{d}} \right\rangle$ is the one-photon Fock state of the detector.

To characterize mechanical entanglement, we adopt the {\em concurrence} $C$ as an entanglement witness \cite{Audenaert2001,Horodecki2009,Wootters1998}, which is traditionally defined for a $(2\times2)$-dimensional mixed state but extends to higher-dimensional states \cite{Audenaert2001}. Maximal entanglement -- as in a pure Bell state -- is characterized by $C=1$, while $C=0$ denotes a fully separable state. The concurrence, computed on the reduced mechanical density operator $\hat \rho_m=\mbox{Tr}_{\hat a,\hat d}[\hat\rho_P]$, is plotted in Fig.\ref{Fig2}(b) as a function of $t_P$. It reaches a maximal value $C>0.9$ at $t_P=83$ ns shortly following the formation of a phonon occupation, highlighted by a vertical dashed line in Figs.\ref{Fig2}(a),(b). $C$ decays as a function of the heralding time $t_P$ according to the mechanical damping rate. In Figs.\ref{Fig2}(c),(d) we plot histograms of the real and imaginary parts of $\hat \rho_m$ restricted to the $\{|00\rangle$, $|01\rangle$, $|10\rangle$, $|11\rangle\}$ basis, as computed at the heralding time where $C$ is maximum. The resemblance to an ideal Bell state emphasizes the nature of the entangled mechanical state. The maximal value of $C$ occurs when $n_c\sim10^{-7}$ \cite{Supplemental}. Given the cavity decay rate, this corresponds to an entanglement heralding rate exceeding $100~\mbox{s}^{-1}$.

\begin{figure}[ht]
\includegraphics[width=0.47\textwidth,clip]{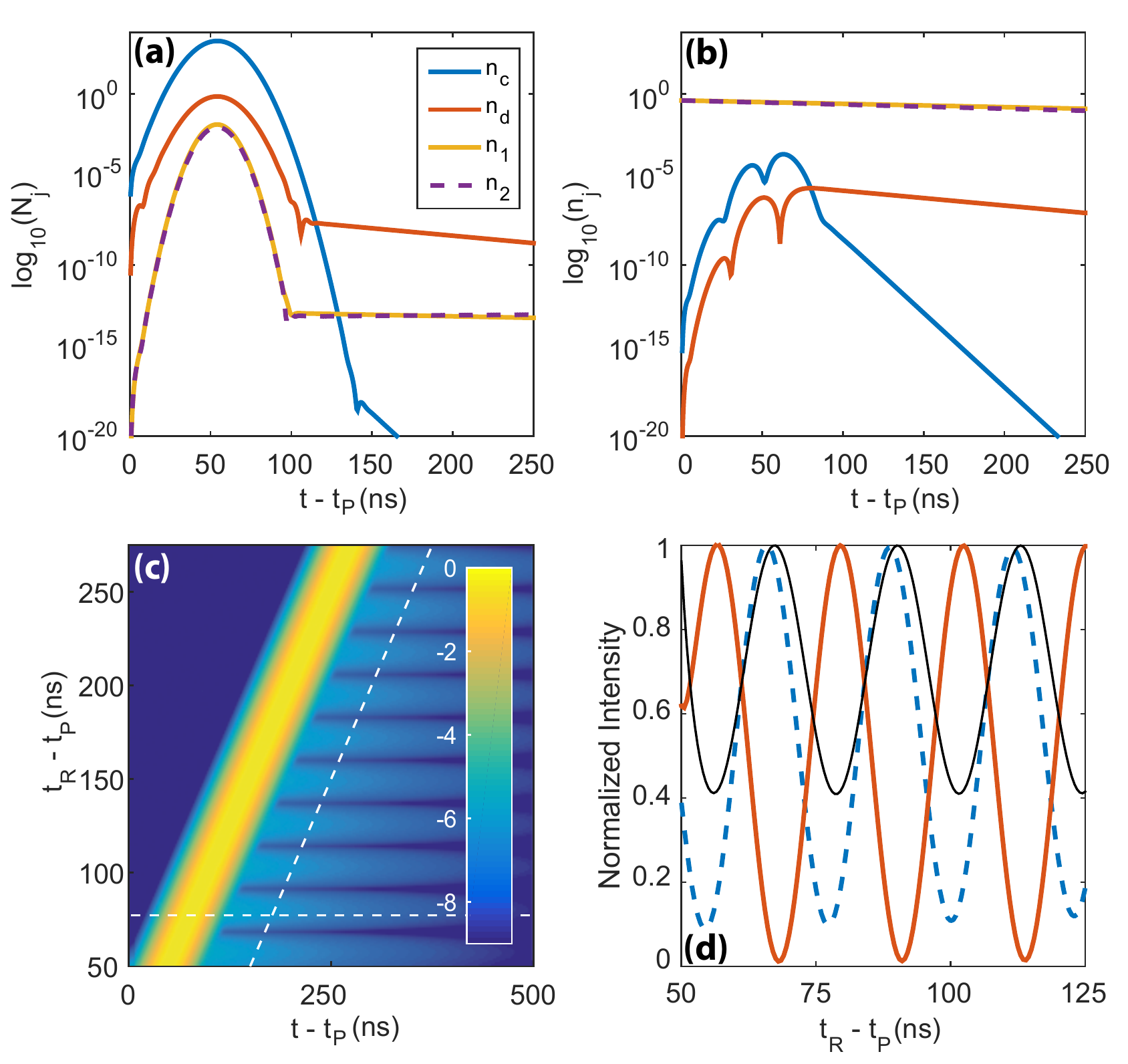}\\
\caption{Readout procedure: (a) Classical and (b) quantum fluctuation contributions to the average occupation of the cavity (blue line), detector (red line), and mechanical (yellow and purple lines) modes. (c) Total intensity $n_d=\langle \hat d^\dagger\hat d\rangle$ at the detector, computed versus the real and readout times. The horizontal dashed line corresponds to the evolution shown in (a) and (b). (d) Interference pattern along the cut denoted by an oblique dashed line in panel (c). The interference pattern at the detector displays the highest visibility ($V =0.97$, red line), as compared to the cavity field ($V =0.82$, blue line) and the detector signal from a fully separable state ($V=0.49$, thin black line).}
\label{Fig3}
\end{figure}

\emph{Readout} --- Readout is accomplished by the second pulse $F_R(t)$. We take $A_R=150\kappa$, $\sigma_R=17.5$ ns, $\omega_R=\omega_c-(\Omega_1+\Omega_2)/2$, and $t_R>t_P$. On a first analysis of the protocol, we set the readout pulse amplitude so that $\langle\hat b_j\rangle\ll1$. This is to avoid the onset of a large, classical mechanical field amplitude that might produce the final interference pattern in the detector even in the absence of entanglement. Figs.\ref{Fig3}(a),(b) show respectively the classical field and the corresponding quantum fluctuation intensities as a function of time for a typical readout process. The classical components to the mechanical fields are negligible compared to the fluctuation part originally created by the heralding. Same holds for the detector field. In Fig.\ref{Fig3}(c), the intensity at the detector is plotted as a function of the real and readout times. After a first strong signal originating from the classical field created in the cavity by the readout pulse, the actual anti-Stokes signal is left, and a clear interference pattern is observed as a function of $t_R$, which is the signature of entanglement between the two mechanical modes. Assuming the pure state $|\psi_2\rangle=\beta_2|0_c\rangle\left(|10\rangle+e^{i\phi(t)}|01\rangle\right)$, i.e. in the ideal case where the ${\left| {00} \right\rangle }$ component vanishes and higher phonon occupations are negligible, the visibility and concurrence are linked via $V=C/(2-C)$. A similar interference pattern is present in the zero delay two-photon correlation $g^{(2)}(0)=\langle\hat d^\dagger\hat d^\dagger\hat d\hat d\rangle/\langle\hat d^\dagger\hat d\rangle^2$ \cite{Supplemental}, highlighting the nonclassicality associated with the interferences. Fig.\ref{Fig3}(d) shows the normalized interference pattern, taken along the oblique dashed line of panel (c). The intensity at the detector (red) has a visibility $V=0.97$, while the intracavity field shows a lower visibility, $V=0.82$, due to the fact that the projection (\ref{rhop}) is carried out at the detector, thus leaving a finite component in the state with $n_c>1$. The black curve shows the interference pattern at the detector when assuming a fully separable state as the initial condition of the readout phase. Its visibility $V\simeq0.5$ sets the lower bound for the existence of entanglement. Fig.\ref{Fig3}(b) shows that the rate of emitted photons after the readout pulse is very low. This, combined with the heralding rate, results in an extremely low success rate for the whole write-readout protocol. This rate would obviously increase with the magnitude of the optomechanical coupling. Another possibility consists in increasing the intensity of both write and readout pulses. For the write pulse, care must be taken to keep the amplitude of the components with two and higher number of mechanical quanta negligible. A stronger readout pulse instead will induce a transient classical field amplitude of the mechanical modes, exceeding quantum fluctuations. Simulations \cite{Supplemental} however show that this amplitude decays rapidly with the readout pulse, and the mechanical Bell state is eventually restored, similarly to recent experimental protocols on micro-macro entanglement \cite{Bruno2013}.

In summary, a single run of the protocol consists in driving the system with the write and readout pulses at given $t_W<t_R$. Only runs where a heralding photon is detected at $t_P<t_R$ are kept. Then the emitted intensity, filtered around $\omega_c$, is then recorded at $t>t_P$. This signal contributes to a horizontal cut of the 3-D plot in Fig. \ref{Fig3}(c). Each run will be characterized by a random heralding time $t_P$. In order to reproduce the data in Fig. \ref{Fig3}(c), the signal should be integrated over several runs to achieve sufficient signal-to-noise ratio. Finally, to maximize entanglement, only runs where $t_P$ falls in a given time window (see Fig. \ref{Fig2}(b)) should be retained.

\emph{Temperature and pure dephasing}.---
We have studied the influence of a finite thermal occupation of the two mechanical modes. Both the visibility $V$ and the concurrence $C$ preserve values exceeding 0.7 up to $\bar{n}_{th}=0.1$, corresponding to a temperature of roughly $T=4~\mbox{mK}$. This temperature might be obtained by a cycle of laser cooling prior to the actual writing step, thanks to the well resolved sideband limit.\cite{Galland2014}

Finally, entanglement can be suppressed by decoherence. Here, we study the effect of pure dephasing acting on the optical degree of freedom, described by a  term $-\eta/2{\cal{D}}\left[ {\hat a^\dag\hat a} \right]\hat \rho$ added to Eq.(\ref{rhot}). The main source of pure dephasing should be optomechanical coupling to other mechanical modes. The pure dephasing rate for a single mechanical mode with frequency $\Omega$, damping rate $\gamma$, and coupling $g$ is estimated by $\eta\sim(g/\Omega)^2(2\bar{n}_{th}+1)\gamma$ \cite{Rabl2011,Supplemental}. Assuming $T=4~\mbox{K}$, and the parameters measured in Ref.\cite{Gavartin2011}, a conservative upper bound to the total pure dephasing rate is $\eta<10^{-7}\kappa$. In this range, the computed values of $C$ and $V$ stay well above 0.5. An analysis of the entanglement as a function of $\bar{n}_{th}$ and $\eta$ is presented in the Supplemental Material \cite{Supplemental}.

\emph{Conclusion} --- We have proposed a realistic protocol for the preparation and readout of mechanical Bell states in a silicon photonic crystal cavity. The cavity quality factor, mechanical mode frequencies, mechanical damping rates and optomechanical coupling strengths all correspond to experimentally demonstrated values for a state-of-the-art system of this kind. The protocol assumes optical cooling of the mechanical modes well within the accessible range, and entanglement is shown to survive the pure dephasing rate expected for this system. The present analysis thus represents a realistic and detailed proposal for an experiment where macroscopic entanglement of two acoustic vibrational modes of a semiconductor microstructure would be produced and detected. Generation of spatially overlapping entangled states at the macroscopic scale holds promise for applications such as quantum state engineering and quantum memories \cite{Lvovsky2009}. An extension of the present analysis to several mechanical modes or multiple write pulses, might indicate the way to the generation of more complex nonclassical states such as GHZ \cite{Bouwmeester1999} or NOON \cite{Kok2002} states.

\begin{acknowledgments}
We are grateful to T. Kippenberg and C. Galland for enlightening discussions.
\end{acknowledgments}

\bibliography{bibliography_arXiv}

\begin{thebibliography}{51}%
\makeatletter
\providecommand \@ifxundefined [1]{%
 \@ifx{#1\undefined}
}%
\providecommand \@ifnum [1]{%
 \ifnum #1\expandafter \@firstoftwo
 \else \expandafter \@secondoftwo
 \fi
}%
\providecommand \@ifx [1]{%
 \ifx #1\expandafter \@firstoftwo
 \else \expandafter \@secondoftwo
 \fi
}%
\providecommand \natexlab [1]{#1}%
\providecommand \enquote  [1]{``#1''}%
\providecommand \bibnamefont  [1]{#1}%
\providecommand \bibfnamefont [1]{#1}%
\providecommand \citenamefont [1]{#1}%
\providecommand \href@noop [0]{\@secondoftwo}%
\providecommand \href [0]{\begingroup \@sanitize@url \@href}%
\providecommand \@href[1]{\@@startlink{#1}\@@href}%
\providecommand \@@href[1]{\endgroup#1\@@endlink}%
\providecommand \@sanitize@url [0]{\catcode `\\12\catcode `\$12\catcode
  `\&12\catcode `\#12\catcode `\^12\catcode `\_12\catcode `\%12\relax}%
\providecommand \@@startlink[1]{}%
\providecommand \@@endlink[0]{}%
\providecommand \url  [0]{\begingroup\@sanitize@url \@url }%
\providecommand \@url [1]{\endgroup\@href {#1}{\urlprefix }}%
\providecommand \urlprefix  [0]{URL }%
\providecommand \Eprint [0]{\href }%
\providecommand \doibase [0]{http://dx.doi.org/}%
\providecommand \selectlanguage [0]{\@gobble}%
\providecommand \bibinfo  [0]{\@secondoftwo}%
\providecommand \bibfield  [0]{\@secondoftwo}%
\providecommand \translation [1]{[#1]}%
\providecommand \BibitemOpen [0]{}%
\providecommand \bibitemStop [0]{}%
\providecommand \bibitemNoStop [0]{.\EOS\space}%
\providecommand \EOS [0]{\spacefactor3000\relax}%
\providecommand \BibitemShut  [1]{\csname bibitem#1\endcsname}%
\let\auto@bib@innerbib\@empty
\bibitem [{\citenamefont {Horodecki}\ \emph {et~al.}(2009)\citenamefont
  {Horodecki}, \citenamefont {Horodecki}, \citenamefont {Horodecki},\ and\
  \citenamefont {Horodecki}}]{Horodecki2009}%
  \BibitemOpen
  \bibfield  {author} {\bibinfo {author} {\bibfnamefont {R.}~\bibnamefont
  {Horodecki}}, \bibinfo {author} {\bibfnamefont {P.}~\bibnamefont
  {Horodecki}}, \bibinfo {author} {\bibfnamefont {M.}~\bibnamefont
  {Horodecki}}, \ and\ \bibinfo {author} {\bibfnamefont {K.}~\bibnamefont
  {Horodecki}},\ }\href {\doibase 10.1103/RevModPhys.81.865} {\bibfield
  {journal} {\bibinfo  {journal} {Rev. Mod. Phys.}\ }\textbf {\bibinfo {volume}
  {81}},\ \bibinfo {pages} {865} (\bibinfo {year} {2009})}\BibitemShut
  {NoStop}%
\bibitem [{\citenamefont {Kimble}(2008)}]{Kimble2008}%
  \BibitemOpen
  \bibfield  {author} {\bibinfo {author} {\bibfnamefont {H.~J.}\ \bibnamefont
  {Kimble}},\ }\href {http://dx.doi.org/10.1038/nature07127} {\bibfield
  {journal} {\bibinfo  {journal} {Nature}\ }\textbf {\bibinfo {volume} {453}},\
  \bibinfo {pages} {1023} (\bibinfo {year} {2008})}\BibitemShut {NoStop}%
\bibitem [{\citenamefont {Duan}\ \emph {et~al.}(2001)\citenamefont {Duan},
  \citenamefont {Lukin}, \citenamefont {Cirac},\ and\ \citenamefont
  {Zoller}}]{Duan2001}%
  \BibitemOpen
  \bibfield  {author} {\bibinfo {author} {\bibfnamefont {L.-M.}\ \bibnamefont
  {Duan}}, \bibinfo {author} {\bibfnamefont {M.~D.}\ \bibnamefont {Lukin}},
  \bibinfo {author} {\bibfnamefont {J.~I.}\ \bibnamefont {Cirac}}, \ and\
  \bibinfo {author} {\bibfnamefont {P.}~\bibnamefont {Zoller}},\ }\href
  {http://dx.doi.org/10.1038/35106500} {\bibfield  {journal} {\bibinfo
  {journal} {Nature}\ }\textbf {\bibinfo {volume} {414}},\ \bibinfo {pages}
  {413} (\bibinfo {year} {2001})}\BibitemShut {NoStop}%
\bibitem [{\citenamefont {Choi}\ \emph {et~al.}(2008)\citenamefont {Choi},
  \citenamefont {Deng}, \citenamefont {Laurat},\ and\ \citenamefont
  {Kimble}}]{Choi2008}%
  \BibitemOpen
  \bibfield  {author} {\bibinfo {author} {\bibfnamefont {K.~S.}\ \bibnamefont
  {Choi}}, \bibinfo {author} {\bibfnamefont {H.}~\bibnamefont {Deng}}, \bibinfo
  {author} {\bibfnamefont {J.}~\bibnamefont {Laurat}}, \ and\ \bibinfo {author}
  {\bibfnamefont {H.~J.}\ \bibnamefont {Kimble}},\ }\href
  {http://dx.doi.org/10.1038/nature06670} {\bibfield  {journal} {\bibinfo
  {journal} {Nature}\ }\textbf {\bibinfo {volume} {452}},\ \bibinfo {pages}
  {67} (\bibinfo {year} {2008})}\BibitemShut {NoStop}%
\bibitem [{\citenamefont {Chou}\ \emph {et~al.}(2005)\citenamefont {Chou},
  \citenamefont {de~Riedmatten}, \citenamefont {Felinto}, \citenamefont
  {Polyakov}, \citenamefont {van Enk},\ and\ \citenamefont
  {Kimble}}]{Chou2005}%
  \BibitemOpen
  \bibfield  {author} {\bibinfo {author} {\bibfnamefont {C.~W.}\ \bibnamefont
  {Chou}}, \bibinfo {author} {\bibfnamefont {H.}~\bibnamefont {de~Riedmatten}},
  \bibinfo {author} {\bibfnamefont {D.}~\bibnamefont {Felinto}}, \bibinfo
  {author} {\bibfnamefont {S.~V.}\ \bibnamefont {Polyakov}}, \bibinfo {author}
  {\bibfnamefont {S.~J.}\ \bibnamefont {van Enk}}, \ and\ \bibinfo {author}
  {\bibfnamefont {H.~J.}\ \bibnamefont {Kimble}},\ }\href
  {http://dx.doi.org/10.1038/nature04353} {\bibfield  {journal} {\bibinfo
  {journal} {Nature}\ }\textbf {\bibinfo {volume} {438}},\ \bibinfo {pages}
  {828} (\bibinfo {year} {2005})}\BibitemShut {NoStop}%
\bibitem [{\citenamefont {Laurat}\ \emph {et~al.}(2007)\citenamefont {Laurat},
  \citenamefont {Choi}, \citenamefont {Deng}, \citenamefont {Chou},\ and\
  \citenamefont {Kimble}}]{Laurat2007}%
  \BibitemOpen
  \bibfield  {author} {\bibinfo {author} {\bibfnamefont {J.}~\bibnamefont
  {Laurat}}, \bibinfo {author} {\bibfnamefont {K.~S.}\ \bibnamefont {Choi}},
  \bibinfo {author} {\bibfnamefont {H.}~\bibnamefont {Deng}}, \bibinfo {author}
  {\bibfnamefont {C.~W.}\ \bibnamefont {Chou}}, \ and\ \bibinfo {author}
  {\bibfnamefont {H.~J.}\ \bibnamefont {Kimble}},\ }\href {\doibase
  10.1103/PhysRevLett.99.180504} {\bibfield  {journal} {\bibinfo  {journal}
  {Phys. Rev. Lett.}\ }\textbf {\bibinfo {volume} {99}},\ \bibinfo {pages}
  {180504} (\bibinfo {year} {2007})}\BibitemShut {NoStop}%
\bibitem [{\citenamefont {Bruno}\ \emph {et~al.}(2013)\citenamefont {Bruno},
  \citenamefont {Martin}, \citenamefont {Sekatski}, \citenamefont {Sangouard},
  \citenamefont {Thew},\ and\ \citenamefont {Gisin}}]{Bruno2013}%
  \BibitemOpen
  \bibfield  {author} {\bibinfo {author} {\bibfnamefont {N.}~\bibnamefont
  {Bruno}}, \bibinfo {author} {\bibfnamefont {A.}~\bibnamefont {Martin}},
  \bibinfo {author} {\bibfnamefont {P.}~\bibnamefont {Sekatski}}, \bibinfo
  {author} {\bibfnamefont {N.}~\bibnamefont {Sangouard}}, \bibinfo {author}
  {\bibfnamefont {R.~T.}\ \bibnamefont {Thew}}, \ and\ \bibinfo {author}
  {\bibfnamefont {N.}~\bibnamefont {Gisin}},\ }\href
  {http://dx.doi.org/10.1038/nphys2681} {\bibfield  {journal} {\bibinfo
  {journal} {Nat Phys}\ }\textbf {\bibinfo {volume} {9}},\ \bibinfo {pages}
  {545} (\bibinfo {year} {2013})}\BibitemShut {NoStop}%
\bibitem [{\citenamefont {Lvovsky}\ \emph {et~al.}(2013)\citenamefont
  {Lvovsky}, \citenamefont {Ghobadi}, \citenamefont {Chandra}, \citenamefont
  {Prasad},\ and\ \citenamefont {Simon}}]{Lvovsky2013}%
  \BibitemOpen
  \bibfield  {author} {\bibinfo {author} {\bibfnamefont {A.~I.}\ \bibnamefont
  {Lvovsky}}, \bibinfo {author} {\bibfnamefont {R.}~\bibnamefont {Ghobadi}},
  \bibinfo {author} {\bibfnamefont {A.}~\bibnamefont {Chandra}}, \bibinfo
  {author} {\bibfnamefont {A.~S.}\ \bibnamefont {Prasad}}, \ and\ \bibinfo
  {author} {\bibfnamefont {C.}~\bibnamefont {Simon}},\ }\href
  {http://dx.doi.org/10.1038/nphys2682} {\bibfield  {journal} {\bibinfo
  {journal} {Nat Phys}\ }\textbf {\bibinfo {volume} {9}},\ \bibinfo {pages}
  {541} (\bibinfo {year} {2013})}\BibitemShut {NoStop}%
\bibitem [{\citenamefont {Usmani}\ \emph {et~al.}(2012)\citenamefont {Usmani},
  \citenamefont {Clausen}, \citenamefont {Bussières}, \citenamefont
  {Sangouard}, \citenamefont {Afzelius},\ and\ \citenamefont
  {Gisin}}]{Usmani2012}%
  \BibitemOpen
  \bibfield  {author} {\bibinfo {author} {\bibfnamefont {I.}~\bibnamefont
  {Usmani}}, \bibinfo {author} {\bibfnamefont {C.}~\bibnamefont {Clausen}},
  \bibinfo {author} {\bibfnamefont {F.}~\bibnamefont {Bussières}}, \bibinfo
  {author} {\bibfnamefont {N.}~\bibnamefont {Sangouard}}, \bibinfo {author}
  {\bibfnamefont {M.}~\bibnamefont {Afzelius}}, \ and\ \bibinfo {author}
  {\bibfnamefont {N.}~\bibnamefont {Gisin}},\ }\href {\doibase
  10.1038/nphoton.2012.34} {\bibfield  {journal} {\bibinfo  {journal} {Nat
  Photon}\ }\textbf {\bibinfo {volume} {6}},\ \bibinfo {pages} {234} (\bibinfo
  {year} {2012})}\BibitemShut {NoStop}%
\bibitem [{\citenamefont {Aspelmeyer}\ \emph {et~al.}(2013)\citenamefont
  {Aspelmeyer}, \citenamefont {Kippenberg},\ and\ \citenamefont
  {Marquardt}}]{Aspelmeyer2013}%
  \BibitemOpen
  \bibfield  {author} {\bibinfo {author} {\bibfnamefont {M.}~\bibnamefont
  {Aspelmeyer}}, \bibinfo {author} {\bibfnamefont {T.~J.}\ \bibnamefont
  {Kippenberg}}, \ and\ \bibinfo {author} {\bibfnamefont {F.}~\bibnamefont
  {Marquardt}},\ }\href {http://arxiv.org/abs/1303.0733} {\bibfield  {journal}
  {\bibinfo  {journal} {{arXiv:1303.0733} [cond-mat, physics:quant-ph]}\ }
  (\bibinfo {year} {2013})}\BibitemShut {NoStop}%
\bibitem [{\citenamefont {Favero}\ and\ \citenamefont
  {Karrai}(2009)}]{Favero2009}%
  \BibitemOpen
  \bibfield  {author} {\bibinfo {author} {\bibfnamefont {I.}~\bibnamefont
  {Favero}}\ and\ \bibinfo {author} {\bibfnamefont {K.}~\bibnamefont
  {Karrai}},\ }\href {http://dx.doi.org/10.1038/nphoton.2009.42} {\bibfield
  {journal} {\bibinfo  {journal} {Nat Photon}\ }\textbf {\bibinfo {volume}
  {3}},\ \bibinfo {pages} {201} (\bibinfo {year} {2009})}\BibitemShut {NoStop}%
\bibitem [{\citenamefont {Kippenberg}\ and\ \citenamefont
  {Vahala}(2008)}]{Kippenberg2008}%
  \BibitemOpen
  \bibfield  {author} {\bibinfo {author} {\bibfnamefont {T.~J.}\ \bibnamefont
  {Kippenberg}}\ and\ \bibinfo {author} {\bibfnamefont {K.~J.}\ \bibnamefont
  {Vahala}},\ }\href {\doibase 10.1126/science.1156032} {\bibfield  {journal}
  {\bibinfo  {journal} {Science}\ }\textbf {\bibinfo {volume} {321}},\ \bibinfo
  {pages} {1172} (\bibinfo {year} {2008})}\BibitemShut {NoStop}%
\bibitem [{\citenamefont {Marquardt}\ and\ \citenamefont
  {Girvin}(2009)}]{Marquardt2009}%
  \BibitemOpen
  \bibfield  {author} {\bibinfo {author} {\bibfnamefont {F.}~\bibnamefont
  {Marquardt}}\ and\ \bibinfo {author} {\bibfnamefont {S.~M.}\ \bibnamefont
  {Girvin}},\ }\href {\doibase 10.1103/Physics.2.40} {\bibfield  {journal}
  {\bibinfo  {journal} {Physics}\ }\textbf {\bibinfo {volume} {2}},\ \bibinfo
  {pages} {40} (\bibinfo {year} {2009})}\BibitemShut {NoStop}%
\bibitem [{\citenamefont {Pirandola}\ \emph {et~al.}(2006)\citenamefont
  {Pirandola}, \citenamefont {Vitali}, \citenamefont {Tombesi},\ and\
  \citenamefont {Lloyd}}]{Pirandola2006}%
  \BibitemOpen
  \bibfield  {author} {\bibinfo {author} {\bibfnamefont {S.}~\bibnamefont
  {Pirandola}}, \bibinfo {author} {\bibfnamefont {D.}~\bibnamefont {Vitali}},
  \bibinfo {author} {\bibfnamefont {P.}~\bibnamefont {Tombesi}}, \ and\
  \bibinfo {author} {\bibfnamefont {S.}~\bibnamefont {Lloyd}},\ }\href
  {\doibase 10.1103/PhysRevLett.97.150403} {\bibfield  {journal} {\bibinfo
  {journal} {Physical Review Letters}\ }\textbf {\bibinfo {volume} {97}},\
  \bibinfo {pages} {150403} (\bibinfo {year} {2006})}\BibitemShut {NoStop}%
\bibitem [{\citenamefont {Zhou}\ \emph {et~al.}(2011)\citenamefont {Zhou},
  \citenamefont {Han}, \citenamefont {Jing},\ and\ \citenamefont
  {Zhang}}]{Zhou2011}%
  \BibitemOpen
  \bibfield  {author} {\bibinfo {author} {\bibfnamefont {L.}~\bibnamefont
  {Zhou}}, \bibinfo {author} {\bibfnamefont {Y.}~\bibnamefont {Han}}, \bibinfo
  {author} {\bibfnamefont {J.}~\bibnamefont {Jing}}, \ and\ \bibinfo {author}
  {\bibfnamefont {W.}~\bibnamefont {Zhang}},\ }\href {\doibase
  10.1103/PhysRevA.83.052117} {\bibfield  {journal} {\bibinfo  {journal}
  {Physical Review A}\ }\textbf {\bibinfo {volume} {83}},\ \bibinfo {pages}
  {052117} (\bibinfo {year} {2011})}\BibitemShut {NoStop}%
\bibitem [{\citenamefont {Vanner}\ \emph {et~al.}(2013)\citenamefont {Vanner},
  \citenamefont {Aspelmeyer},\ and\ \citenamefont {Kim}}]{Vanner2013}%
  \BibitemOpen
  \bibfield  {author} {\bibinfo {author} {\bibfnamefont {M.~R.}\ \bibnamefont
  {Vanner}}, \bibinfo {author} {\bibfnamefont {M.}~\bibnamefont {Aspelmeyer}},
  \ and\ \bibinfo {author} {\bibfnamefont {M.~S.}\ \bibnamefont {Kim}},\ }\href
  {\doibase 10.1103/PhysRevLett.110.010504} {\bibfield  {journal} {\bibinfo
  {journal} {Physical Review Letters}\ }\textbf {\bibinfo {volume} {110}},\
  \bibinfo {pages} {010504} (\bibinfo {year} {2013})}\BibitemShut {NoStop}%
\bibitem [{\citenamefont {Ghobadi}\ \emph {et~al.}(2014)\citenamefont
  {Ghobadi}, \citenamefont {Kumar}, \citenamefont {Pepper}, \citenamefont
  {Bouwmeester}, \citenamefont {Lvovsky},\ and\ \citenamefont
  {Simon}}]{Ghobadi2014}%
  \BibitemOpen
  \bibfield  {author} {\bibinfo {author} {\bibfnamefont {R.}~\bibnamefont
  {Ghobadi}}, \bibinfo {author} {\bibfnamefont {S.}~\bibnamefont {Kumar}},
  \bibinfo {author} {\bibfnamefont {B.}~\bibnamefont {Pepper}}, \bibinfo
  {author} {\bibfnamefont {D.}~\bibnamefont {Bouwmeester}}, \bibinfo {author}
  {\bibfnamefont {A.}~\bibnamefont {Lvovsky}}, \ and\ \bibinfo {author}
  {\bibfnamefont {C.}~\bibnamefont {Simon}},\ }\href {\doibase
  10.1103/PhysRevLett.112.080503} {\bibfield  {journal} {\bibinfo  {journal}
  {Physical Review Letters}\ }\textbf {\bibinfo {volume} {112}},\ \bibinfo
  {pages} {080503} (\bibinfo {year} {2014})}\BibitemShut {NoStop}%
\bibitem [{\citenamefont {Galland}\ \emph {et~al.}(2014)\citenamefont
  {Galland}, \citenamefont {Sangouard}, \citenamefont {Piro}, \citenamefont
  {Gisin},\ and\ \citenamefont {Kippenberg}}]{Galland2014}%
  \BibitemOpen
  \bibfield  {author} {\bibinfo {author} {\bibfnamefont {C.}~\bibnamefont
  {Galland}}, \bibinfo {author} {\bibfnamefont {N.}~\bibnamefont {Sangouard}},
  \bibinfo {author} {\bibfnamefont {N.}~\bibnamefont {Piro}}, \bibinfo {author}
  {\bibfnamefont {N.}~\bibnamefont {Gisin}}, \ and\ \bibinfo {author}
  {\bibfnamefont {T.~J.}\ \bibnamefont {Kippenberg}},\ }\href {\doibase
  10.1103/PhysRevLett.112.143602} {\bibfield  {journal} {\bibinfo  {journal}
  {Physical Review Letters}\ }\textbf {\bibinfo {volume} {112}},\ \bibinfo
  {pages} {143602} (\bibinfo {year} {2014})}\BibitemShut {NoStop}%
\bibitem [{\citenamefont {Akram}\ \emph {et~al.}(2013)\citenamefont {Akram},
  \citenamefont {Bowen},\ and\ \citenamefont {Milburn}}]{Akram2013}%
  \BibitemOpen
  \bibfield  {author} {\bibinfo {author} {\bibfnamefont {U.}~\bibnamefont
  {Akram}}, \bibinfo {author} {\bibfnamefont {W.~P.}\ \bibnamefont {Bowen}}, \
  and\ \bibinfo {author} {\bibfnamefont {G.~J.}\ \bibnamefont {Milburn}},\
  }\href {\doibase 10.1088/1367-2630/15/9/093007} {\bibfield  {journal}
  {\bibinfo  {journal} {New Journal of Physics}\ }\textbf {\bibinfo {volume}
  {15}},\ \bibinfo {pages} {093007} (\bibinfo {year} {2013})}\BibitemShut
  {NoStop}%
\bibitem [{\citenamefont {Borkje}\ \emph {et~al.}(2011)\citenamefont {Borkje},
  \citenamefont {Nunnenkamp},\ and\ \citenamefont {Girvin}}]{Borkje2011}%
  \BibitemOpen
  \bibfield  {author} {\bibinfo {author} {\bibfnamefont {K.}~\bibnamefont
  {Borkje}}, \bibinfo {author} {\bibfnamefont {A.}~\bibnamefont {Nunnenkamp}},
  \ and\ \bibinfo {author} {\bibfnamefont {S.~M.}\ \bibnamefont {Girvin}},\
  }\href {\doibase 10.1103/PhysRevLett.107.123601} {\bibfield  {journal}
  {\bibinfo  {journal} {Physical Review Letters}\ }\textbf {\bibinfo {volume}
  {107}},\ \bibinfo {pages} {123601} (\bibinfo {year} {2011})}\BibitemShut
  {NoStop}%
\bibitem [{\citenamefont {Ge}\ \emph {et~al.}(2013)\citenamefont {Ge},
  \citenamefont {Al-Amri}, \citenamefont {Nha},\ and\ \citenamefont
  {Zubairy}}]{Ge2013}%
  \BibitemOpen
  \bibfield  {author} {\bibinfo {author} {\bibfnamefont {W.}~\bibnamefont
  {Ge}}, \bibinfo {author} {\bibfnamefont {M.}~\bibnamefont {Al-Amri}},
  \bibinfo {author} {\bibfnamefont {H.}~\bibnamefont {Nha}}, \ and\ \bibinfo
  {author} {\bibfnamefont {M.~S.}\ \bibnamefont {Zubairy}},\ }\href {\doibase
  10.1103/PhysRevA.88.022338} {\bibfield  {journal} {\bibinfo  {journal}
  {Physical Review A}\ }\textbf {\bibinfo {volume} {88}},\ \bibinfo {pages}
  {022338} (\bibinfo {year} {2013})}\BibitemShut {NoStop}%
\bibitem [{\citenamefont {Hartmann}\ and\ \citenamefont
  {Plenio}(2008)}]{Hartmann2008}%
  \BibitemOpen
  \bibfield  {author} {\bibinfo {author} {\bibfnamefont {M.~J.}\ \bibnamefont
  {Hartmann}}\ and\ \bibinfo {author} {\bibfnamefont {M.~B.}\ \bibnamefont
  {Plenio}},\ }\href {\doibase 10.1103/PhysRevLett.101.200503} {\bibfield
  {journal} {\bibinfo  {journal} {Physical Review Letters}\ }\textbf {\bibinfo
  {volume} {101}},\ \bibinfo {pages} {200503} (\bibinfo {year}
  {2008})}\BibitemShut {NoStop}%
\bibitem [{\citenamefont {Ludwig}\ \emph {et~al.}(2010)\citenamefont {Ludwig},
  \citenamefont {Hammerer},\ and\ \citenamefont {Marquardt}}]{Ludwig2010}%
  \BibitemOpen
  \bibfield  {author} {\bibinfo {author} {\bibfnamefont {M.}~\bibnamefont
  {Ludwig}}, \bibinfo {author} {\bibfnamefont {K.}~\bibnamefont {Hammerer}}, \
  and\ \bibinfo {author} {\bibfnamefont {F.}~\bibnamefont {Marquardt}},\ }\href
  {\doibase 10.1103/PhysRevA.82.012333} {\bibfield  {journal} {\bibinfo
  {journal} {Physical Review A}\ }\textbf {\bibinfo {volume} {82}},\ \bibinfo
  {pages} {012333} (\bibinfo {year} {2010})}\BibitemShut {NoStop}%
\bibitem [{\citenamefont {Pinard}\ \emph {et~al.}(2005)\citenamefont {Pinard},
  \citenamefont {Dantan}, \citenamefont {Vitali}, \citenamefont {Arcizet},
  \citenamefont {Briant},\ and\ \citenamefont {Heidmann}}]{Pinard2005}%
  \BibitemOpen
  \bibfield  {author} {\bibinfo {author} {\bibfnamefont {M.}~\bibnamefont
  {Pinard}}, \bibinfo {author} {\bibfnamefont {A.}~\bibnamefont {Dantan}},
  \bibinfo {author} {\bibfnamefont {D.}~\bibnamefont {Vitali}}, \bibinfo
  {author} {\bibfnamefont {O.}~\bibnamefont {Arcizet}}, \bibinfo {author}
  {\bibfnamefont {T.}~\bibnamefont {Briant}}, \ and\ \bibinfo {author}
  {\bibfnamefont {A.}~\bibnamefont {Heidmann}},\ }\href {\doibase
  10.1209/epl/i2005-10317-6} {\bibfield  {journal} {\bibinfo  {journal} {{EPL}
  (Europhysics Letters)}\ }\textbf {\bibinfo {volume} {72}},\ \bibinfo {pages}
  {747} (\bibinfo {year} {2005})}\BibitemShut {NoStop}%
\bibitem [{\citenamefont {Szorkovszky}\ \emph {et~al.}(2014)\citenamefont
  {Szorkovszky}, \citenamefont {Clerk}, \citenamefont {Doherty},\ and\
  \citenamefont {Bowen}}]{Szorkovszky2014}%
  \BibitemOpen
  \bibfield  {author} {\bibinfo {author} {\bibfnamefont {A.}~\bibnamefont
  {Szorkovszky}}, \bibinfo {author} {\bibfnamefont {A.~A.}\ \bibnamefont
  {Clerk}}, \bibinfo {author} {\bibfnamefont {A.~C.}\ \bibnamefont {Doherty}},
  \ and\ \bibinfo {author} {\bibfnamefont {W.~P.}\ \bibnamefont {Bowen}},\
  }\href {http://arxiv.org/abs/1402.6392} {\bibfield  {journal} {\bibinfo
  {journal} {{arXiv:1402.6392} [cond-mat, physics:quant-ph]}\ } (\bibinfo
  {year} {2014})}\BibitemShut {NoStop}%
\bibitem [{\citenamefont {Tan}\ \emph {et~al.}(2013)\citenamefont {Tan},
  \citenamefont {Li},\ and\ \citenamefont {Meystre}}]{Tan2013a}%
  \BibitemOpen
  \bibfield  {author} {\bibinfo {author} {\bibfnamefont {H.}~\bibnamefont
  {Tan}}, \bibinfo {author} {\bibfnamefont {G.}~\bibnamefont {Li}}, \ and\
  \bibinfo {author} {\bibfnamefont {P.}~\bibnamefont {Meystre}},\ }\href
  {\doibase 10.1103/PhysRevA.87.033829} {\bibfield  {journal} {\bibinfo
  {journal} {Physical Review A}\ }\textbf {\bibinfo {volume} {87}},\ \bibinfo
  {pages} {033829} (\bibinfo {year} {2013})}\BibitemShut {NoStop}%
\bibitem [{\citenamefont {Xu}\ \emph {et~al.}(2013)\citenamefont {Xu},
  \citenamefont {Zhao},\ and\ \citenamefont {Liu}}]{Xu2013}%
  \BibitemOpen
  \bibfield  {author} {\bibinfo {author} {\bibfnamefont {X.-W.}\ \bibnamefont
  {Xu}}, \bibinfo {author} {\bibfnamefont {Y.-J.}\ \bibnamefont {Zhao}}, \ and\
  \bibinfo {author} {\bibfnamefont {Y.-x.}\ \bibnamefont {Liu}},\ }\href
  {\doibase 10.1103/PhysRevA.88.022325} {\bibfield  {journal} {\bibinfo
  {journal} {Physical Review A}\ }\textbf {\bibinfo {volume} {88}},\ \bibinfo
  {pages} {022325} (\bibinfo {year} {2013})}\BibitemShut {NoStop}%
\bibitem [{\citenamefont {Lee}\ \emph {et~al.}(2011)\citenamefont {Lee},
  \citenamefont {Sprague}, \citenamefont {Sussman}, \citenamefont {Nunn},
  \citenamefont {Langford}, \citenamefont {Jin}, \citenamefont {Champion},
  \citenamefont {Michelberger}, \citenamefont {Reim}, \citenamefont {England},
  \citenamefont {Jaksch},\ and\ \citenamefont {Walmsley}}]{Lee2011}%
  \BibitemOpen
  \bibfield  {author} {\bibinfo {author} {\bibfnamefont {K.~C.}\ \bibnamefont
  {Lee}}, \bibinfo {author} {\bibfnamefont {M.~R.}\ \bibnamefont {Sprague}},
  \bibinfo {author} {\bibfnamefont {B.~J.}\ \bibnamefont {Sussman}}, \bibinfo
  {author} {\bibfnamefont {J.}~\bibnamefont {Nunn}}, \bibinfo {author}
  {\bibfnamefont {N.~K.}\ \bibnamefont {Langford}}, \bibinfo {author}
  {\bibfnamefont {X.-M.}\ \bibnamefont {Jin}}, \bibinfo {author} {\bibfnamefont
  {T.}~\bibnamefont {Champion}}, \bibinfo {author} {\bibfnamefont
  {P.}~\bibnamefont {Michelberger}}, \bibinfo {author} {\bibfnamefont {K.~F.}\
  \bibnamefont {Reim}}, \bibinfo {author} {\bibfnamefont {D.}~\bibnamefont
  {England}}, \bibinfo {author} {\bibfnamefont {D.}~\bibnamefont {Jaksch}}, \
  and\ \bibinfo {author} {\bibfnamefont {I.~A.}\ \bibnamefont {Walmsley}},\
  }\href {\doibase 10.1126/science.1211914} {\bibfield  {journal} {\bibinfo
  {journal} {Science}\ }\textbf {\bibinfo {volume} {334}},\ \bibinfo {pages}
  {1253} (\bibinfo {year} {2011})},\ \bibinfo {note} {{PMID:}
  22144620}\BibitemShut {NoStop}%
\bibitem [{\citenamefont {Gavartin}\ \emph {et~al.}(2011)\citenamefont
  {Gavartin}, \citenamefont {Braive}, \citenamefont {Sagnes}, \citenamefont
  {Arcizet}, \citenamefont {Beveratos}, \citenamefont {Kippenberg},\ and\
  \citenamefont {Robert-Philip}}]{Gavartin2011}%
  \BibitemOpen
  \bibfield  {author} {\bibinfo {author} {\bibfnamefont {E.}~\bibnamefont
  {Gavartin}}, \bibinfo {author} {\bibfnamefont {R.}~\bibnamefont {Braive}},
  \bibinfo {author} {\bibfnamefont {I.}~\bibnamefont {Sagnes}}, \bibinfo
  {author} {\bibfnamefont {O.}~\bibnamefont {Arcizet}}, \bibinfo {author}
  {\bibfnamefont {A.}~\bibnamefont {Beveratos}}, \bibinfo {author}
  {\bibfnamefont {T.~J.}\ \bibnamefont {Kippenberg}}, \ and\ \bibinfo {author}
  {\bibfnamefont {I.}~\bibnamefont {Robert-Philip}},\ }\href {\doibase
  10.1103/PhysRevLett.106.203902} {\bibfield  {journal} {\bibinfo  {journal}
  {Physical Review Letters}\ }\textbf {\bibinfo {volume} {106}},\ \bibinfo
  {pages} {203902} (\bibinfo {year} {2011})}\BibitemShut {NoStop}%
\bibitem [{\citenamefont {Ding}\ \emph {et~al.}(2011)\citenamefont {Ding},
  \citenamefont {Baker}, \citenamefont {Senellart}, \citenamefont {Lemaitre},
  \citenamefont {Ducci}, \citenamefont {Leo},\ and\ \citenamefont
  {Favero}}]{Ding2011}%
  \BibitemOpen
  \bibfield  {author} {\bibinfo {author} {\bibfnamefont {L.}~\bibnamefont
  {Ding}}, \bibinfo {author} {\bibfnamefont {C.}~\bibnamefont {Baker}},
  \bibinfo {author} {\bibfnamefont {P.}~\bibnamefont {Senellart}}, \bibinfo
  {author} {\bibfnamefont {A.}~\bibnamefont {Lemaitre}}, \bibinfo {author}
  {\bibfnamefont {S.}~\bibnamefont {Ducci}}, \bibinfo {author} {\bibfnamefont
  {G.}~\bibnamefont {Leo}}, \ and\ \bibinfo {author} {\bibfnamefont
  {I.}~\bibnamefont {Favero}},\ }\href {\doibase
  http://dx.doi.org/10.1063/1.3563711} {\bibfield  {journal} {\bibinfo
  {journal} {Applied Physics Letters}\ }\textbf {\bibinfo {volume} {98}},\
  \bibinfo {eid} {113108} (\bibinfo {year} {2011})}\BibitemShut {NoStop}%
\bibitem [{\citenamefont {Chan}\ \emph {et~al.}(2011)\citenamefont {Chan},
  \citenamefont {Alegre}, \citenamefont {Safavi-Naeini}, \citenamefont {Hill},
  \citenamefont {Krause}, \citenamefont {Groblacher}, \citenamefont
  {Aspelmeyer},\ and\ \citenamefont {Painter}}]{Chan2011}%
  \BibitemOpen
  \bibfield  {author} {\bibinfo {author} {\bibfnamefont {J.}~\bibnamefont
  {Chan}}, \bibinfo {author} {\bibfnamefont {T.~P.~M.}\ \bibnamefont {Alegre}},
  \bibinfo {author} {\bibfnamefont {A.~H.}\ \bibnamefont {Safavi-Naeini}},
  \bibinfo {author} {\bibfnamefont {J.~T.}\ \bibnamefont {Hill}}, \bibinfo
  {author} {\bibfnamefont {A.}~\bibnamefont {Krause}}, \bibinfo {author}
  {\bibfnamefont {S.}~\bibnamefont {Groblacher}}, \bibinfo {author}
  {\bibfnamefont {M.}~\bibnamefont {Aspelmeyer}}, \ and\ \bibinfo {author}
  {\bibfnamefont {O.}~\bibnamefont {Painter}},\ }\href
  {http://dx.doi.org/10.1038/nature10461} {\bibfield  {journal} {\bibinfo
  {journal} {Nature}\ }\textbf {\bibinfo {volume} {478}},\ \bibinfo {pages}
  {89} (\bibinfo {year} {2011})}\BibitemShut {NoStop}%
\bibitem [{\citenamefont {Chan}\ \emph {et~al.}(2012)\citenamefont {Chan},
  \citenamefont {Safavi-Naeini}, \citenamefont {Hill}, \citenamefont
  {Meenehan},\ and\ \citenamefont {Painter}}]{Chan2012}%
  \BibitemOpen
  \bibfield  {author} {\bibinfo {author} {\bibfnamefont {J.}~\bibnamefont
  {Chan}}, \bibinfo {author} {\bibfnamefont {A.~H.}\ \bibnamefont
  {Safavi-Naeini}}, \bibinfo {author} {\bibfnamefont {J.~T.}\ \bibnamefont
  {Hill}}, \bibinfo {author} {\bibfnamefont {S.}~\bibnamefont {Meenehan}}, \
  and\ \bibinfo {author} {\bibfnamefont {O.}~\bibnamefont {Painter}},\ }\href
  {\doibase http://dx.doi.org/10.1063/1.4747726} {\bibfield  {journal}
  {\bibinfo  {journal} {Applied Physics Letters}\ }\textbf {\bibinfo {volume}
  {101}},\ \bibinfo {eid} {081115} (\bibinfo {year} {2012})}\BibitemShut
  {NoStop}%
\bibitem [{\citenamefont {Gavartin}\ \emph {et~al.}(2013)\citenamefont
  {Gavartin}, \citenamefont {Verlot},\ and\ \citenamefont
  {Kippenberg}}]{Gavartin2013}%
  \BibitemOpen
  \bibfield  {author} {\bibinfo {author} {\bibfnamefont {E.}~\bibnamefont
  {Gavartin}}, \bibinfo {author} {\bibfnamefont {P.}~\bibnamefont {Verlot}}, \
  and\ \bibinfo {author} {\bibfnamefont {T.~J.}\ \bibnamefont {Kippenberg}},\
  }\href {http://dx.doi.org/10.1038/ncomms3860} {\bibfield  {journal} {\bibinfo
   {journal} {Nat Commun}\ }\textbf {\bibinfo {volume} {4}},\  (\bibinfo {year}
  {2013})}\BibitemShut {NoStop}%
\bibitem [{\citenamefont {Nguyen}\ \emph {et~al.}(2013)\citenamefont {Nguyen},
  \citenamefont {Baker}, \citenamefont {Hease}, \citenamefont {Sejil},
  \citenamefont {Senellart}, \citenamefont {Lemaître}, \citenamefont {Ducci},
  \citenamefont {Leo},\ and\ \citenamefont {Favero}}]{Nguyen2013}%
  \BibitemOpen
  \bibfield  {author} {\bibinfo {author} {\bibfnamefont {D.~T.}\ \bibnamefont
  {Nguyen}}, \bibinfo {author} {\bibfnamefont {C.}~\bibnamefont {Baker}},
  \bibinfo {author} {\bibfnamefont {W.}~\bibnamefont {Hease}}, \bibinfo
  {author} {\bibfnamefont {S.}~\bibnamefont {Sejil}}, \bibinfo {author}
  {\bibfnamefont {P.}~\bibnamefont {Senellart}}, \bibinfo {author}
  {\bibfnamefont {A.}~\bibnamefont {Lemaître}}, \bibinfo {author}
  {\bibfnamefont {S.}~\bibnamefont {Ducci}}, \bibinfo {author} {\bibfnamefont
  {G.}~\bibnamefont {Leo}}, \ and\ \bibinfo {author} {\bibfnamefont
  {I.}~\bibnamefont {Favero}},\ }\href {\doibase
  http://dx.doi.org/10.1063/1.4846515} {\bibfield  {journal} {\bibinfo
  {journal} {Applied Physics Letters}\ }\textbf {\bibinfo {volume} {103}},\
  \bibinfo {eid} {241112} (\bibinfo {year} {2013})}\BibitemShut {NoStop}%
\bibitem [{\citenamefont {Safavi-Naeini}\ \emph {et~al.}(2013)\citenamefont
  {Safavi-Naeini}, \citenamefont {Groblacher}, \citenamefont {Hill},
  \citenamefont {Chan}, \citenamefont {Aspelmeyer},\ and\ \citenamefont
  {Painter}}]{Safavi-Naeini2013}%
  \BibitemOpen
  \bibfield  {author} {\bibinfo {author} {\bibfnamefont {A.~H.}\ \bibnamefont
  {Safavi-Naeini}}, \bibinfo {author} {\bibfnamefont {S.}~\bibnamefont
  {Groblacher}}, \bibinfo {author} {\bibfnamefont {J.~T.}\ \bibnamefont
  {Hill}}, \bibinfo {author} {\bibfnamefont {J.}~\bibnamefont {Chan}}, \bibinfo
  {author} {\bibfnamefont {M.}~\bibnamefont {Aspelmeyer}}, \ and\ \bibinfo
  {author} {\bibfnamefont {O.}~\bibnamefont {Painter}},\ }\href
  {http://dx.doi.org/10.1038/nature12307} {\bibfield  {journal} {\bibinfo
  {journal} {Nature}\ }\textbf {\bibinfo {volume} {500}},\ \bibinfo {pages}
  {185} (\bibinfo {year} {2013})}\BibitemShut {NoStop}%
\bibitem [{\citenamefont {Safavi-Naeini}\ \emph {et~al.}(2014)\citenamefont
  {Safavi-Naeini}, \citenamefont {Hill}, \citenamefont {Meenehan},
  \citenamefont {Chan}, \citenamefont {Gr\"oblacher},\ and\ \citenamefont
  {Painter}}]{Safavi-Naeini2014}%
  \BibitemOpen
  \bibfield  {author} {\bibinfo {author} {\bibfnamefont {A.~H.}\ \bibnamefont
  {Safavi-Naeini}}, \bibinfo {author} {\bibfnamefont {J.~T.}\ \bibnamefont
  {Hill}}, \bibinfo {author} {\bibfnamefont {S.}~\bibnamefont {Meenehan}},
  \bibinfo {author} {\bibfnamefont {J.}~\bibnamefont {Chan}}, \bibinfo {author}
  {\bibfnamefont {S.}~\bibnamefont {Gr\"oblacher}}, \ and\ \bibinfo {author}
  {\bibfnamefont {O.}~\bibnamefont {Painter}},\ }\href {\doibase
  10.1103/PhysRevLett.112.153603} {\bibfield  {journal} {\bibinfo  {journal}
  {Phys. Rev. Lett.}\ }\textbf {\bibinfo {volume} {112}},\ \bibinfo {pages}
  {153603} (\bibinfo {year} {2014})}\BibitemShut {NoStop}%
\bibitem [{\citenamefont {Safavi-Naeini}\ and\ \citenamefont
  {Painter}(2010)}]{Safavi-Naeini2010}%
  \BibitemOpen
  \bibfield  {author} {\bibinfo {author} {\bibfnamefont {A.~H.}\ \bibnamefont
  {Safavi-Naeini}}\ and\ \bibinfo {author} {\bibfnamefont {O.}~\bibnamefont
  {Painter}},\ }\href {\doibase 10.1364/OE.18.014926} {\bibfield  {journal}
  {\bibinfo  {journal} {Opt. Express}\ }\textbf {\bibinfo {volume} {18}},\
  \bibinfo {pages} {14926} (\bibinfo {year} {2010})}\BibitemShut {NoStop}%
\bibitem [{\citenamefont {Lai}\ \emph {et~al.}(2014)\citenamefont {Lai},
  \citenamefont {Pirotta}, \citenamefont {Urbinati}, \citenamefont {Gerace},
  \citenamefont {Minkov}, \citenamefont {Savona}, \citenamefont {Badolato},\
  and\ \citenamefont {Galli}}]{Lai2014}%
  \BibitemOpen
  \bibfield  {author} {\bibinfo {author} {\bibfnamefont {Y.}~\bibnamefont
  {Lai}}, \bibinfo {author} {\bibfnamefont {S.}~\bibnamefont {Pirotta}},
  \bibinfo {author} {\bibfnamefont {G.}~\bibnamefont {Urbinati}}, \bibinfo
  {author} {\bibfnamefont {D.}~\bibnamefont {Gerace}}, \bibinfo {author}
  {\bibfnamefont {M.}~\bibnamefont {Minkov}}, \bibinfo {author} {\bibfnamefont
  {V.}~\bibnamefont {Savona}}, \bibinfo {author} {\bibfnamefont
  {A.}~\bibnamefont {Badolato}}, \ and\ \bibinfo {author} {\bibfnamefont
  {M.}~\bibnamefont {Galli}},\ }\href {\doibase
  http://dx.doi.org/10.1063/1.4882860} {\bibfield  {journal} {\bibinfo
  {journal} {Applied Physics Letters}\ }\textbf {\bibinfo {volume} {104}},\
  \bibinfo {eid} {241101} (\bibinfo {year} {2014})}\BibitemShut {NoStop}%
\bibitem [{\citenamefont {Minkov}\ and\ \citenamefont
  {Savona}(2014)}]{Minkov2014}%
  \BibitemOpen
  \bibfield  {author} {\bibinfo {author} {\bibfnamefont {M.}~\bibnamefont
  {Minkov}}\ and\ \bibinfo {author} {\bibfnamefont {V.}~\bibnamefont
  {Savona}},\ }\href {http://dx.doi.org/10.1038/srep05124} {\bibfield
  {journal} {\bibinfo  {journal} {Sci. Rep.}\ }\textbf {\bibinfo {volume}
  {4}},\  (\bibinfo {year} {2014})}\BibitemShut {NoStop}%
\bibitem [{Sup()}]{Supplemental}%
  \BibitemOpen
  \href@noop {} {\bibinfo  {journal} {See Supplemental Material below for more
  details}\ }\BibitemShut {NoStop}%
\bibitem [{\citenamefont {Mandel}(1999)}]{Mandel1999}%
  \BibitemOpen
\bibfield  {journal} {  }\bibfield  {author} {\bibinfo {author} {\bibfnamefont
  {L.}~\bibnamefont {Mandel}},\ }\href {\doibase 10.1103/RevModPhys.71.S274}
  {\bibfield  {journal} {\bibinfo  {journal} {Rev. Mod. Phys.}\ }\textbf
  {\bibinfo {volume} {71}},\ \bibinfo {pages} {S274} (\bibinfo {year}
  {1999})}\BibitemShut {NoStop}%
\bibitem [{\citenamefont {Zou}\ \emph {et~al.}(1991)\citenamefont {Zou},
  \citenamefont {Wang},\ and\ \citenamefont {Mandel}}]{Zou1991}%
  \BibitemOpen
  \bibfield  {author} {\bibinfo {author} {\bibfnamefont {X.~Y.}\ \bibnamefont
  {Zou}}, \bibinfo {author} {\bibfnamefont {L.~J.}\ \bibnamefont {Wang}}, \
  and\ \bibinfo {author} {\bibfnamefont {L.}~\bibnamefont {Mandel}},\ }\href
  {\doibase 10.1103/PhysRevLett.67.318} {\bibfield  {journal} {\bibinfo
  {journal} {Phys. Rev. Lett.}\ }\textbf {\bibinfo {volume} {67}},\ \bibinfo
  {pages} {318} (\bibinfo {year} {1991})}\BibitemShut {NoStop}%
\bibitem [{\citenamefont {del Valle}\ \emph {et~al.}(2012)\citenamefont {del
  Valle}, \citenamefont {Gonzalez-Tudela}, \citenamefont {Laussy},
  \citenamefont {Tejedor},\ and\ \citenamefont {Hartmann}}]{Valle2012}%
  \BibitemOpen
  \bibfield  {author} {\bibinfo {author} {\bibfnamefont {E.}~\bibnamefont {del
  Valle}}, \bibinfo {author} {\bibfnamefont {A.}~\bibnamefont
  {Gonzalez-Tudela}}, \bibinfo {author} {\bibfnamefont {F.~P.}\ \bibnamefont
  {Laussy}}, \bibinfo {author} {\bibfnamefont {C.}~\bibnamefont {Tejedor}}, \
  and\ \bibinfo {author} {\bibfnamefont {M.~J.}\ \bibnamefont {Hartmann}},\
  }\href {\doibase 10.1103/PhysRevLett.109.183601} {\bibfield  {journal}
  {\bibinfo  {journal} {Phys. Rev. Lett.}\ }\textbf {\bibinfo {volume} {109}},\
  \bibinfo {pages} {183601} (\bibinfo {year} {2012})}\BibitemShut {NoStop}%
\bibitem [{\citenamefont {Gardiner}\ and\ \citenamefont
  {Zoller}(2004)}]{Gardiner2004}%
  \BibitemOpen
  \bibfield  {author} {\bibinfo {author} {\bibfnamefont {C.}~\bibnamefont
  {Gardiner}}\ and\ \bibinfo {author} {\bibfnamefont {P.}~\bibnamefont
  {Zoller}},\ }\href {http://books.google.ch/books?id=a\_xsT8oGhdgC} {\emph
  {\bibinfo {title} {Quantum Noise: A Handbook of Markovian and Non-Markovian
  Quantum Stochastic Methods with Applications to Quantum Optics}}},\ Springer
  Series in Synergetics\ (\bibinfo  {publisher} {Springer},\ \bibinfo {year}
  {2004})\BibitemShut {NoStop}%
\bibitem [{\citenamefont {Audenaert}\ \emph {et~al.}(2001)\citenamefont
  {Audenaert}, \citenamefont {Verstraete},\ and\ \citenamefont
  {De~Moor}}]{Audenaert2001}%
  \BibitemOpen
  \bibfield  {author} {\bibinfo {author} {\bibfnamefont {K.}~\bibnamefont
  {Audenaert}}, \bibinfo {author} {\bibfnamefont {F.}~\bibnamefont
  {Verstraete}}, \ and\ \bibinfo {author} {\bibfnamefont {B.}~\bibnamefont
  {De~Moor}},\ }\href {\doibase 10.1103/PhysRevA.64.052304} {\bibfield
  {journal} {\bibinfo  {journal} {Phys. Rev. A}\ }\textbf {\bibinfo {volume}
  {64}},\ \bibinfo {pages} {052304} (\bibinfo {year} {2001})}\BibitemShut
  {NoStop}%
\bibitem [{\citenamefont {Wootters}(1998)}]{Wootters1998}%
  \BibitemOpen
  \bibfield  {author} {\bibinfo {author} {\bibfnamefont {W.~K.}\ \bibnamefont
  {Wootters}},\ }\href {\doibase 10.1103/PhysRevLett.80.2245} {\bibfield
  {journal} {\bibinfo  {journal} {Phys. Rev. Lett.}\ }\textbf {\bibinfo
  {volume} {80}},\ \bibinfo {pages} {2245} (\bibinfo {year}
  {1998})}\BibitemShut {NoStop}%
\bibitem [{\citenamefont {Rabl}(2011)}]{Rabl2011}%
  \BibitemOpen
  \bibfield  {author} {\bibinfo {author} {\bibfnamefont {P.}~\bibnamefont
  {Rabl}},\ }\href {\doibase 10.1103/PhysRevLett.107.063601} {\bibfield
  {journal} {\bibinfo  {journal} {Phys. Rev. Lett.}\ }\textbf {\bibinfo
  {volume} {107}},\ \bibinfo {pages} {063601} (\bibinfo {year}
  {2011})}\BibitemShut {NoStop}%
\bibitem [{\citenamefont {Lvovsky}\ \emph {et~al.}(2009)\citenamefont
  {Lvovsky}, \citenamefont {Sanders},\ and\ \citenamefont
  {Tittel}}]{Lvovsky2009}%
  \BibitemOpen
  \bibfield  {author} {\bibinfo {author} {\bibfnamefont {A.~I.}\ \bibnamefont
  {Lvovsky}}, \bibinfo {author} {\bibfnamefont {B.~C.}\ \bibnamefont
  {Sanders}}, \ and\ \bibinfo {author} {\bibfnamefont {W.}~\bibnamefont
  {Tittel}},\ }\href {\doibase 10.1038/nphoton.2009.231} {\bibfield  {journal}
  {\bibinfo  {journal} {Nature Photonics}\ }\textbf {\bibinfo {volume} {3}},\
  \bibinfo {pages} {706} (\bibinfo {year} {2009})}\BibitemShut {NoStop}%
\bibitem [{\citenamefont {Bouwmeester}\ \emph {et~al.}(1999)\citenamefont
  {Bouwmeester}, \citenamefont {Pan}, \citenamefont {Daniell}, \citenamefont
  {Weinfurter},\ and\ \citenamefont {Zeilinger}}]{Bouwmeester1999}%
  \BibitemOpen
  \bibfield  {author} {\bibinfo {author} {\bibfnamefont {D.}~\bibnamefont
  {Bouwmeester}}, \bibinfo {author} {\bibfnamefont {J.-W.}\ \bibnamefont
  {Pan}}, \bibinfo {author} {\bibfnamefont {M.}~\bibnamefont {Daniell}},
  \bibinfo {author} {\bibfnamefont {H.}~\bibnamefont {Weinfurter}}, \ and\
  \bibinfo {author} {\bibfnamefont {A.}~\bibnamefont {Zeilinger}},\ }\href
  {\doibase 10.1103/PhysRevLett.82.1345} {\bibfield  {journal} {\bibinfo
  {journal} {Phys. Rev. Lett.}\ }\textbf {\bibinfo {volume} {82}},\ \bibinfo
  {pages} {1345} (\bibinfo {year} {1999})}\BibitemShut {NoStop}%
\bibitem [{\citenamefont {Kok}\ \emph {et~al.}(2002)\citenamefont {Kok},
  \citenamefont {Lee},\ and\ \citenamefont {Dowling}}]{Kok2002}%
  \BibitemOpen
  \bibfield  {author} {\bibinfo {author} {\bibfnamefont {P.}~\bibnamefont
  {Kok}}, \bibinfo {author} {\bibfnamefont {H.}~\bibnamefont {Lee}}, \ and\
  \bibinfo {author} {\bibfnamefont {J.~P.}\ \bibnamefont {Dowling}},\ }\href
  {\doibase 10.1103/PhysRevA.65.052104} {\bibfield  {journal} {\bibinfo
  {journal} {Phys. Rev. A}\ }\textbf {\bibinfo {volume} {65}},\ \bibinfo
  {pages} {052104} (\bibinfo {year} {2002})}\BibitemShut {NoStop}%
\bibitem [{\citenamefont {Nunnenkamp}\ \emph {et~al.}(2011)\citenamefont
  {Nunnenkamp}, \citenamefont {Børkje},\ and\ \citenamefont
  {Girvin}}]{Nunnenkamp2011}%
  \BibitemOpen
  \bibfield  {author} {\bibinfo {author} {\bibfnamefont {A.}~\bibnamefont
  {Nunnenkamp}}, \bibinfo {author} {\bibfnamefont {K.}~\bibnamefont {Børkje}},
  \ and\ \bibinfo {author} {\bibfnamefont {S.~M.}\ \bibnamefont {Girvin}},\
  }\href {\doibase 10.1103/PhysRevLett.107.063602} {\bibfield  {journal}
  {\bibinfo  {journal} {Physical Review Letters}\ }\textbf {\bibinfo {volume}
  {107}},\ \bibinfo {pages} {063602} (\bibinfo {year} {2011})}\BibitemShut
  {NoStop}%
\end{thebibliography}%

\pagebreak
\begin{widetext}
\begin{center}
\textbf{\large Supplemental Material}
\end{center}
We detail here the semiclassical model deployed for readout procedure. We provide some additional information on the second order correlation function during readout, demonstrating the single photon emission, and on the impact of temperature and pure dephasing. We discuss the potential improvement of the single photon emission rates. Finally we provide some analytical expressions linking the fringes visibility to the mechanical concurrence.
\end{widetext}
\setcounter{equation}{0}
\setcounter{figure}{0}

\section{Readout Semiclassical Model}
The readout procedure was modeled by distinguishing the quantum fluctuations $\delta \hat a$, $\delta \hat{d}$ and $\delta \hat b_j$ from the classical field amplitudes $\alpha=\langle {\hat a}\rangle$, $\delta=\langle {\hat d} \rangle$ and $\beta_j=\langle {\hat{b}_j}\rangle$. More specifically, the total fields are expressed as $\hat a = \alpha + \delta \hat a$, $\hat d = \delta + \delta \hat d$ and $\hat b_j = \beta_j + \delta \hat b_j$. This procedure allows modeling the effect of a large amplitude of the driving field, as assumed in the readout phase. An intense driving field typically induces large average occupations of the modes under study. Of these, only a small part is contributed from quantum fluctuations, while the largest part is accounted for by the classical-field-component, similarly to the textbook case of a displaced quantum oscillator. Hence, the separation into the two contributions makes the numerical analysis possible while still restricting the Hilbert space to a reasonably small number of fluctuation quanta. In such a displaced representation, where the classical fields act as the new vacuum, the system Hamiltonian reads
\begin{equation}\label{H}
{\cal{\hat H}} = \sum\limits_{j = 1,2} \left[\begin{array}{l} 2{g_j}{\mathop{\rm Re}\nolimits} \left( {{\beta _j}} \right){{\hat a}^\dag }{\hat a} + \hbar{\Omega_j}\hat b_j^\dag {{\hat b}_j}\\
+ {g_j}\left( {{\alpha ^*}{{\hat a}} + \alpha \hat a^\dag } \right)\left( {\hat b_j^\dag  + {{\hat b}_j}} \right)\\
+ {g_j}\hat a^\dag {{\hat a}}\left( {\hat b_j^\dag  + {{\hat b}_j}} \right) \end{array}\right]
\end{equation}
where we have simplified the notations as $\delta \hat{o} \rightarrow \hat{o}$. Note that contrary to the usual linearisation procedure we keep here terms of all orders in ${\cal{\hat H}}$ \cite{Nunnenkamp2011}. The quantum fluctuations dynamics is governed by the master equation
\begin{eqnarray}
\nonumber \frac{{d\hat \rho }}{{dt}} = &-& i\left[ {{\cal{\hat H}},\hat \rho } \right] - \frac{{{\kappa}}}{2}{\cal{D}}\left[ {\hat a} \right]\hat \rho  - \frac{{{\kappa _d}}}{2}{\cal{D}}\left[ {\hat d} \right]\hat \rho + \frac{{{\zeta }}}{2}{{\cal{D}}}\left[ \hat a, \hat d \right]\hat \rho \\
\label{rhot}
 &-& {{\bar n}_{th}}\sum\limits_j {\frac{\gamma_j }{2} {{\cal{D}}\left[ {{{\hat b}_j}} \right]\hat \rho  - \left( {{{\bar n}_{th}} + 1} \right){\cal{D}}\left[ {\hat b_j^\dag } \right]\hat \rho }}\,.
\end{eqnarray}
that is coupled the time-evolution equations for the classical field components
\begin{eqnarray}
i\dot \alpha  &=&  - i\frac{\kappa }{2}\alpha  + 2\alpha\sum\limits_{j = 1,2}  {g_j}{\mathop{\rm Re}\nolimits} \left( {{\beta _j}} \right) + F\left( t \right)\\
i{{\dot \beta }_j} &=& \left( {\hbar{\Omega _j} - i\frac{{{\gamma _j}}}{2}} \right){\beta _j} + {g_j}{\left| \alpha  \right|^2}\\
i\dot \delta  &=&  - i\frac{{{\kappa _d}}}{2}\delta  + i\frac{\zeta }{2}\alpha
\end{eqnarray}
We assume the projected mechanical density matrix $\hat{\rho}_m$, occurring in the write phase at the time where the concurrence is maximal [see Fig.2(b)], as the initial condition for the dynamics in the readout phase. In this framework, the total average occupation of the cavity mode for example is evaluated as $n_c = \langle {{{\hat a}^\dag }\hat a} \rangle + {\left| \alpha  \right|^2}$, and similar expressions hold for the other modes.

\section{Second Order Correlations}
We show in Fig.S1(a),(b) the second order correlation functions at zero delay versus time of the cavity and detector fields, respectively defined as
\begin{equation}
g_c^{\left( 2 \right)}\left( t,t \right) = \frac{{\langle {{{\hat a}^\dag }{{\hat a}^\dag }\hat a\hat a} \rangle }}{{{{\langle {{{\hat a}^\dag }\hat a} \rangle }^2}}},~~~g_d^{\left( 2 \right)}\left(t,t \right) = \frac{{\langle {{{\hat d}^\dag }{{\hat d}^\dag }\hat d\hat d}\rangle }}{{{{\langle {{{\hat d}^\dag }\hat d}\rangle }^2}}}
\end{equation}
along the oblique cut highlighted in Fig.3(c). Here, the operators encompass the total classical and fluctuation contributions. We have superimposed the normalized intensity (blue line) and we clearly see the strong antibunching obtained at the intensity maximum. This antibunching is the signature of the emission of single photons as a result of the anti-Stokes raman process that destroys the mechanical Bell state as a result of the readout pulse.

\begin{figure}[ht!]
\renewcommand{\figurename}{Fig.S}
\includegraphics[width=0.49\textwidth,clip]{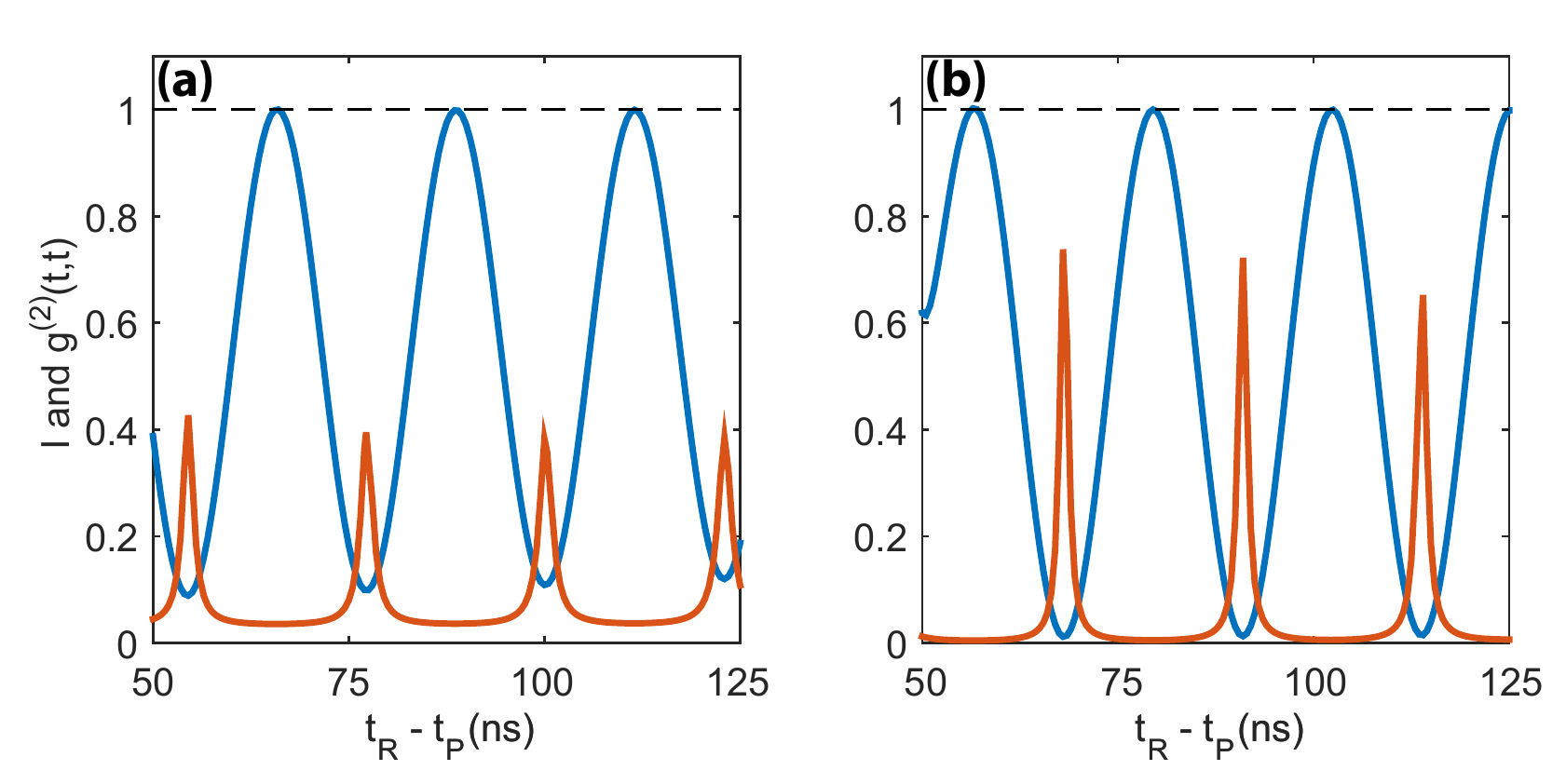}\\
\caption{(Color online). Second order correlation at zero delay (red curves) and normalised intensity $I_S$ (blue curves) along the oblique cut of Fig.3(c) for (a) the cavity and (b) the detector fields.}
\label{FigS1}
\end{figure}

\section{Impact of temperature and pure dephasing}
We have analyzed the impact of a finite temperature on the concurrence of the entangled mechanical states and on the fringes visibility. The results are presented in Fig.S2(a). We obtain $C\simeq0.7$ at $\bar{n}_{th}=0.1$,  which correspond to typical temperatures attainable with sideband cooling \cite{Galland2014}. We clearly see the correlation between the visibility and the concurrence versus $\bar{n}_{th}$. Indeed, the contribution of the mechanical Fock states with more that one quantum increases with temperature. It results in a larger matrix element of the $|11\rangle\langle 11|$ component of ${\hat \rho }_m$ which, as suggested in the main text, reduces entanglement and the fringe visibility.

Finally, in Ref.\cite{Gavartin2011}, at least four extra mechanical modes were shown to exhibit a non-negligible optomechanical coupling to the cavity. These modes form sidebands that are spectrally unresolved from the main cavity line, and induce pure dephasing of the cavity mode. This can be taken into account adding a pure dephasing Linblad term $-\eta/2{\cal{D}}\left[ {\hat a^\dag\hat a} \right]\hat \rho$ to the master equation. An estimation of $\eta$ can be made \cite{Rabl2011}, accounting for the collective effect of all other mechanical modes, according to
\begin{equation}\label{xi}
\eta  = \sum\limits_j {\left( {{{\bar n }_{j,th}} + 1} \right){\gamma _j}\frac{{g_j^2}}{{\Omega _j^2}}}
\end{equation}
where $j$ runs over all extra mechanical modes and $\bar n_{j,th}=1/[\exp(\hbar \Omega_j/k_B T)-1]$. Assuming an average mechanical quality factor of $Q=400$ and the frequency and optomechanical coupling parameters measured in Ref. \cite{Gavartin2011}, we obtain an upper boundary as small as $\eta\simeq10^{-7}$ for a bath temperature of $T=1$K. The impact on the concurrence and visibility are reported in Fig.S2(b) for $\eta$ spanning several decades. We see that the concurrence drops faster than the fringes visibility. Indeed, pure dephasing mostly affects the photon/phonon correlations and therefore the projection efficiency during the write procedure. It translates into an increase in the $|00\rangle\langle 00|$ element of the reduced mechanical density matrix ${\hat \rho }_{m}$ after the heralding.

\begin{figure}[ht]
\renewcommand{\figurename}{Fig.S}
\includegraphics[width=0.49\textwidth,clip]{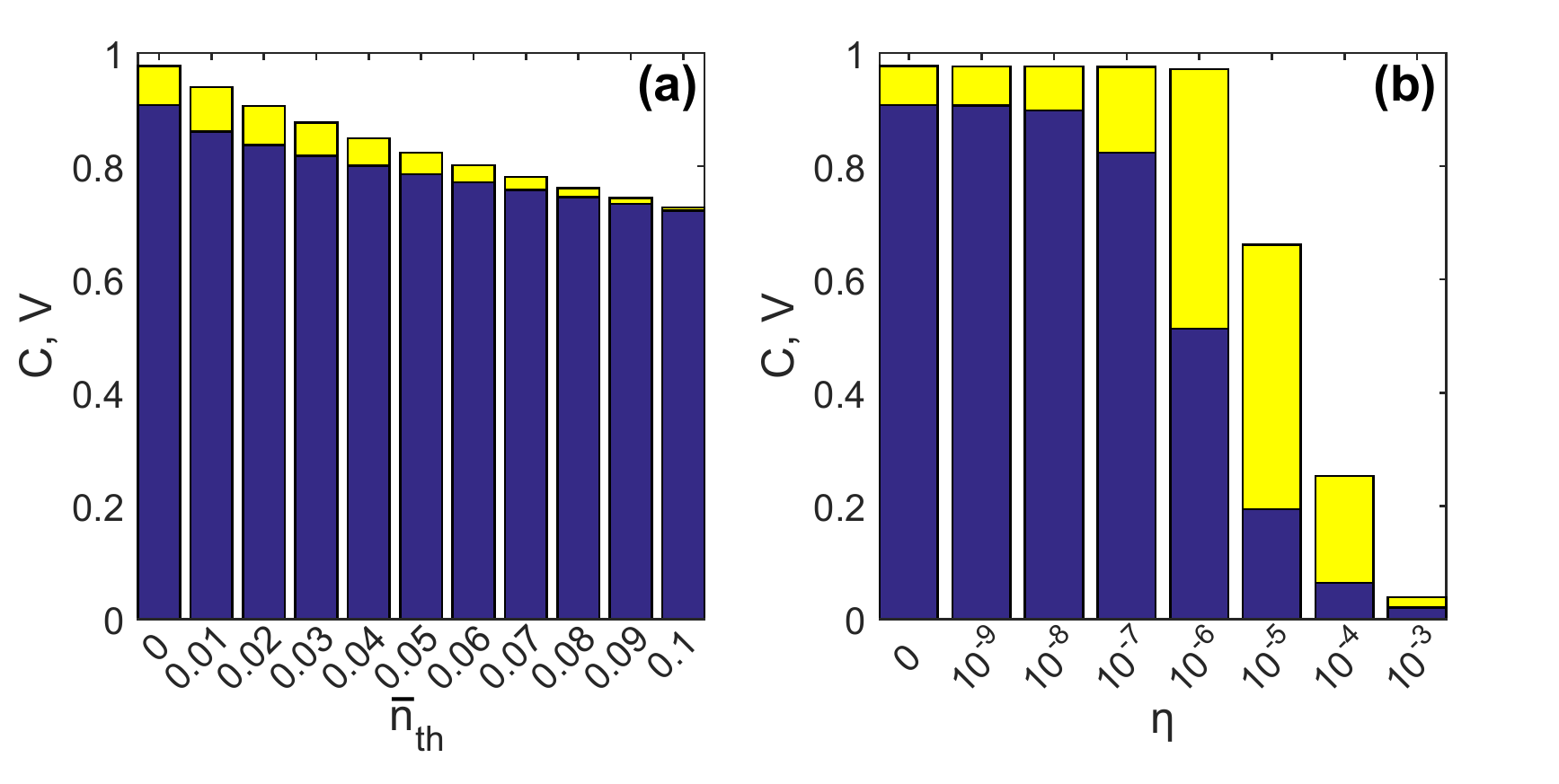}\\
\caption{(Color online). Impact of finite thermal phonon occupation (a) and pure dephasing rate (b) on the concurrence $C$ (blue bars) and fringe visibility $V$ (yellow bars).}
\label{FigS2}
\end{figure}

\section{Emission rates improvement}

Here, we discuss possible ways to increase the rate of successful heralding/readout events in the protocol. The discussion is developed separately for the write and readout phases:

\emph{Write} --- In Fig.2(a),(b) we have seen that during the write procedure, the maximum concurrence is obtained at a time where the average cavity occupation is of the order of $10^{-7}$, which in turn sets the heralding efficiency. The heralding rate could be improved with a larger optomechanical coupling (which would induce a faster cavity/mechanical correlation time) or by increasing the intensity of the write pulse. In the latter case, one has to keep in mind that a limit is set by the requirement that the density matrix elements relative to states with two or more mechanical quanta be negligible compared to the elements relative to states with zero or one quantum. The parameters used in our simulations correspond to a situation largely within this boundary, and therefore leave considerable room for improvement.

A third possibility is to compromise to a lower value of the concurrence. As seen in Figs.2(a) and (b) of the main manuscript, if for example a concurrence $C=0.1$ is accepted, such a value occurs at earlier heralding times, where the average cavity occupation is about $2\times10^{-3}$. This is highlighted in the Fig.S3(a) showing the cavity occupation $n_c(t)$ versus $C(t)$.

\emph{Readout} --- In Fig.3(a),(b) of the main manuscript, the emission rate of single readout photons in the region where the fluctuations are dominating, turns out to be of the order of $10^{-9}$, i.e. corresponding to an exceedingly low readout rate when combined to the heralding rate. The associated parameters where chosen in order to keep the classical contribution to the phonon fields negligible. We have however investigated larger pump amplitude regimes where the classical contribution to the phonon fields dominates over the fluctuations. In Figs.S3(b), (c), and (d) we have used a pump amplitude of $A_R = 5000\kappa$ while all the other parameters where left unchanged. The interference fringes shown in panel (b) are still marked and the fluctuation contribution to the average occupations (panel (d)) reaches $2\times10^{-4}$. We have verified that, while a large classical component to the mechanical fields develops (see panel (c)), the net effect of such component is simply to displace the mechanical Bell state temporarily. Once the readout pulse has decayed, this classical component also vanishes and the mechanical mode evolves back into the Bell state. Note finally that, for such strong readout field intensity, the system approaches the strong optomechanical coupling regime, corresponding to $g_j\alpha\sim\kappa$, and mechanical backaction on the cavity field becomes sizeable, as one can see in the panel (d).

\begin{figure}[ht!]
\renewcommand{\figurename}{Fig.S}
\includegraphics[width=0.49\textwidth,clip]{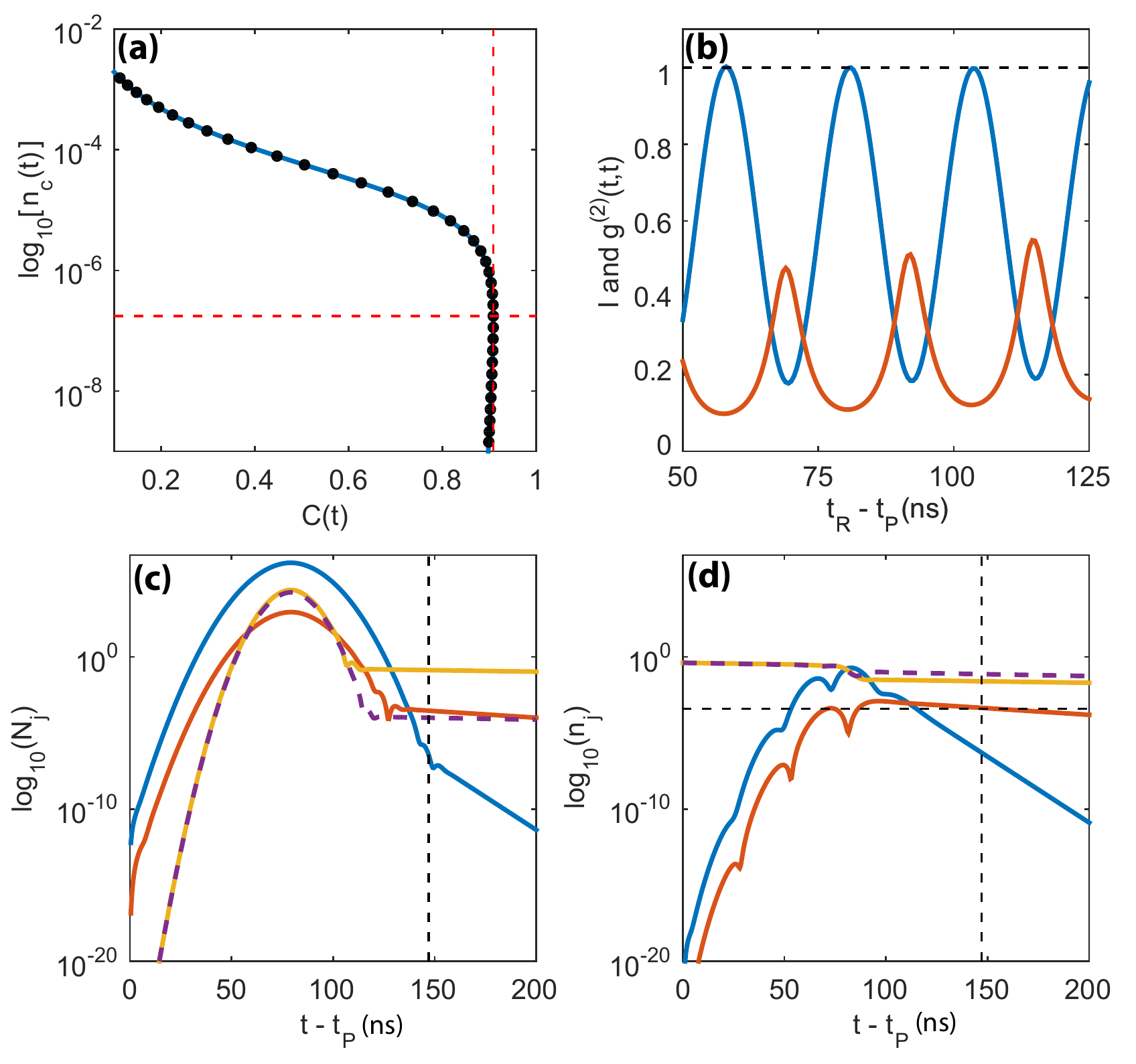}\\
\caption{(a) Average cavity occupation versus concurrence corresponding to the evolution of Fig.2(a),(b) of the main text. The dashed lines highlight the $C$ maximum. Classical (c) and fluctuation (d) contributions to the field intensity during the readout procedure for $A_R = 5000\kappa$. (b) Corresponding interference fringes (blue curve $V=0.7$) and second order correlation at zero delay of the detector mode.}
\label{FigS3}
\end{figure}

\section{Visibility and Concurrence}
We derive here analytical expressions linking the visibility to the concurrence, in the simple case where a pure state is assumed.

\emph{Visibility}.---
Let us consider the following state characterizing the system immediately after the write procedure (i.e. after the heralding photon has been detected)
\begin{equation}
\left| {{\psi _W}} \right\rangle  = \left| {{0_c}} \right\rangle \left( {{c_{00}}\left| {00} \right\rangle  + {c_{10}}\left| {10} \right\rangle  + {c_{01}}\left| {01} \right\rangle  + {c_{11}}\left| {11} \right\rangle } \right)
\end{equation}
where $c_{ij}$ are complex coefficients fulfilling the normalization condition ${\left| {{c_{00}}} \right|^2} + {\left| {{c_{01}}} \right|^2} + {\left| {{c_{10}}} \right|^2} + {\left| {{c_{11}}} \right|^2} = 1$ . The readout procedure based on anti-Stokes processes ideally corresponds to the time-evolution induced by the Hamiltonian ${\hat{\cal{{H}}}}_R = {{\hat a}^\dag }\left( {{{\hat b}_1} + {{\hat b}_2}} \right) + \hat a\left( {\hat b_1^\dag  + \hat b_2^\dag } \right)$, which eventually results in the state
\begin{eqnarray}
\nonumber \left| {{\psi _R}} \right\rangle  &=& \left| {{0_c}} \right\rangle \left[ {{c_{00}}\left| {10} \right\rangle  + {c_{00}}\left| {01} \right\rangle  + \left( {{c_{10}} + {c_{01}}} \right)\left| {11} \right\rangle } \right]\\
 &+& \left| {{1_c}} \right\rangle \left[ {{c_{11}}\left| {10} \right\rangle  + {c_{11}}\left| {01} \right\rangle + \left( {{c_{10}} + {c_{01}}} \right)\left| {00} \right\rangle } \right]
\end{eqnarray}
The corresponding cavity mode intensity is given by
\begin{eqnarray}
\nonumber {I_R} &=& \left\langle {{\psi _R}} \right|{{\hat a}^\dag }\hat a\left| {{\psi _R}} \right\rangle \\
 &=& {\left| {{c_{10}} + {c_{01}}} \right|^2} + 2{\left| {{c_{11}}} \right|^2}
\end{eqnarray}
Consequently the maximally entangled $\left| {{\psi _B}} \right\rangle  = \left( {\left| {10} \right\rangle  + {e^{i\phi \left( t_R \right)}}\left| {01} \right\rangle } \right)/\sqrt 2$ and the fully separable $\left| {{\psi _S}} \right\rangle  = \left( {\left| 0 \right\rangle  + \left| 1 \right\rangle } \right)\left( {\left| 0 \right\rangle  + {e^{i\phi \left( t_R \right)}}\left| 1 \right\rangle } \right)/\sqrt 2$ states produce cavity intensities
\begin{eqnarray}
{I_B} &=& 1 + \cos \left[ {\phi \left( {{t_R}} \right)} \right]\\
{I_S} &=& 1 + \cos \left[ {\phi \left( {{t_R}} \right)} \right]/2
\end{eqnarray}
respectively as a function of the readout time $t_R$. The corresponding visibilities are therefore $V_B=1$ and $V_S=0.5$.

\emph{Concurrence}.---
For the two mode system defined by the state $\left| {{\psi _W}} \right\rangle$ the associated density matrix $\hat{\rho}_m  = \left| \psi  \right\rangle \left\langle \psi  \right|$ reads
\begin{equation}
{\hat \rho} = \left( {\begin{array}{*{20}{c}}
{{{\left| {{c_{00}}} \right|}^2}}&{{c_{00}}c_{01}^*}&{{c_{00}}c_{10}^*}&{{c_{00}}c_{11}^*}\\
{{c_{01}}c_{00}^*}&{{{\left| {{c_{01}}} \right|}^2}}&{{c_{01}}c_{10}^*}&{{c_{01}}c_{11}^*}\\
{{c_{10}}c_{00}^*}&{{c_{10}}c_{01}^*}&{{{\left| {{c_{10}}} \right|}^2}}&{{c_{10}}c_{11}^*}\\
{{c_{11}}c_{00}^*}&{{c_{11}}c_{01}^*}&{{c_{11}}c_{10}^*}&{{{\left| {{c_{11}}} \right|}^2}}
\end{array}} \right)
\end{equation}
The concurrence \cite{Wootters1998} is defined as $C\left({{\hat{\rho}_m}}\right) = \max \left( {0,{\lambda _1} - {\lambda _2} - {\lambda _3} - {\lambda _4}} \right)$ where $\lambda_j$ are square roots of the eigenvalues in decreasing order of the non-Hermitian matrix ${\cal{R}}=\hat{\rho}_m{\tilde{\rho}}_m$ where
\begin{equation}
{{\tilde \rho_m }} = \left( {{\sigma _y} \otimes {\sigma _y}} \right)\hat{\rho}_m^*\left( {{\sigma _y} \otimes {\sigma _y}} \right)
\end{equation}
is the so-called spin-flipped density matrix. It turns out that, in the case considered here, only a single eigenvalue of ${{\tilde \rho_m }}$ is nonzero and we obtain
\begin{equation}
C\left( {{\hat{\rho}_m}}\right) = 2\left|  {{c_{10}}c_{01} - {c_{00}}c_{11}}  \right|
\end{equation}
One can easily check that for example $\left| {{\psi _B}} \right\rangle$ is characterized by $C=1$ while $\left| {{\psi _S}} \right\rangle$ gives $C=0$.

More generally assuming that $c_{01}=c_{10}e^{i \phi}$, the intensity reduces to $I = 2{\left| {{c_{10}}} \right|^2}\left( {\cos \phi  - 1} \right) + 2\left| {{c_{11}}} \right|^2$ resulting in a visibility
\begin{equation}
V = \frac{{{{\left| {{c_{10}}} \right|}^2}}}{{{{\left| {{c_{10}}} \right|}^2} + {{\left| {{c_{11}}} \right|}^2}}}
\end{equation}
From this expression it is immediately seen that $V=0$ for $c_{10}=0$, $V=1$ for $c_{11}=0$. More generally, in order to have $V>0.5$, one needs ${\left| {{c_{10}}} \right|^2}, {\left| {{c_{01}}} \right|^2}> {\left| {{c_{11}}} \right|^2}$. Note however that although the coefficient $c_{00}$ doesn't enter directly the intensity and visibility expressions it is linked to the other coefficients via the normalization condition ${\left| {{c_{00}}} \right|^2} = 1 - 2{\left| {{c_{10}}} \right|^2} - {\left| {{c_{11}}} \right|^2}$.

In the protocol that we propose, immediately after the heralding photon has been detected, the vacuum mechanical state has been projected out and we can assume $c_{00}=0$. Then the visibility, with the help of the normalization relation for the remaining coefficients, is given by $V = {\left| {{c_{10}}} \right|^2}/(1 - {\left| {{c_{10}}} \right|^2})$, and can be directly expressed as a function of the concurrence $C=2{\left| {{c_{10}}} \right|^2}$ namely
\begin{eqnarray}
V = \frac{C}{{2 - C}}
\end{eqnarray}


\end{document}